# Comprehensive QM/ML Study of Dislocation Loop Line Segments in α-Zr Part I: ⟨a⟩ Loop Core Geometries, Solute Binding, and Transition Barriers

Junting Zhang[1], Andrew P. Horsfield[2] and Mark R. Wenman[1]

1. Department of Materials and Centre for Nuclear Engineering, Imperial College London, London, UK
2. Department of Materials and Thomas Young Centre, Imperial College London, UK

Corresponding author: Mark R. Wenman
Corresponding author email: m.wenman@imperial.ac.uk

## Abstract:

Zr cladding alloys in nuclear reactors are exposed to high neutron fluxes, making irradiation-induced defects and solute segregation central to understanding their in-service behaviour. Dislocation cores involve short-range chemistry and long-range elastic fields, which are difficult to describe simultaneously using conventional atomistic methods. Using a hybrid quantum mechanics/machine learning (QM/ML) framework, this part of work examines prismatic ⟨a⟩ loops in hcp α-Zr. All three loop variants are considered, including faulted and perfect, interstitial- and vacancy-type, and first- and second-order prismatic configurations. Nine possible dislocation line segments are characterised in terms of core geometry, solute binding, glide barriers and faulted-to-perfect transition pathways using QM/ML and newly implemented QM/ML-NEB calculations. Shockley and half-Shockley partials, together with basal and prismatic fault ribbons, are found to govern core stabilisation. Fe and H preferentially bind to faulted segments, whereas Sn and Nb favour perfect segments. Solutes can also drive partial motion, recombination and local unfaulting. These results establish a comprehensive picture of ⟨a⟩ loop microstructure, solute trapping and transition mechanisms in α-Zr.

# 1. Introduction

Zr cladding alloys are key structural materials in light-water reactor fuel assemblies because they combine good mechanical performance, corrosion resistance, and a low thermal-neutron capture cross section [1, 2, 3]. Their in-reactor degradation is mainly governed by irradiation-induced mechanical damage and corrosion-assisted hydrogen uptake [4, 5, 6]. Under neutron irradiation, displacement cascades generate large populations of point defects that migrate, interact, and collapse into extended defects such as dislocation loops [4, 5]. These loops drive microstructural evolution, irradiation hardening, and dimensional instability, commonly referred to as mechanical degradation and irradiation-induced growth [7, 8]. In addition to their mechanical effects, dislocation loops also influence alloy chemistry: atom probe tomography and transmission electron microscopy studies show that they can act as sinks for solutes and impurities like Sn, Nb, Fe and H, promoting radiation-induced segregation and planar solute clustering after irradiation [9, 10].

Dislocation loops in the hcp α-Zr matrix are generally classified into two major groups according to their Burgers vector and habit plane: ⟨a⟩ loops and ⟨c⟩-component loops. Prismatic slip is more readily activated than basal or pyramidal slip in α-Zr. Irradiation-induced loops formed at lower doses are predominantly prismatic ⟨a⟩ loops [7, 11, 12]. These loops may be vacancy or interstitial in character and can potentially glide on prismatic planes, making them particularly important for irradiation hardening and plastic deformation [5, 13, 14]. At higher irradiation doses, basal-plane ⟨c⟩-component loops can also form; these are commonly associated with vacancy character and are widely regarded as a major contributor to irradiation-induced growth [7, 8]. Among the possible loop variants in irradiated α-Zr, ⟨a⟩ loops are therefore central to mechanical degradation, whereas ⟨c⟩-component loops are especially important for dimensional instability [7, 8, 15, 16]. This paper focuses on ⟨a⟩ loops and systematically examines th eir possible core and line-segment variants.

More specifically, atomistic studies have shown that, beyond their vacancy or interstitial character, ⟨a⟩ loops in α-Zr can be further divided into three subclasses according to their habit plane and Burgers vector, as illustrated in Fig. 1 [12, 16]. In hcp crystals, prismatic planes are grouped into first-order prismatic planes, $\{10\bar{1}0\}$, and second-order prismatic planes, $\{11\bar{2}0\}$, with most reported ⟨a⟩ loops lying on the first-order prismatic plane. The first type is the faulted first-order prismatic ⟨a⟩ loop (Fig. 1a), which has Burgers vector $1/2\langle 10\bar{1}0\rangle$. Because this Burgers vector is normal to the loop habit plane, the loop is pure edge in character. In α-Zr, the $\langle 10\bar{1}0\rangle$ directions, follows an ABCDABCD stacking sequence, and collapse of vacancy or interstitial clusters into such a loop produces a metastable prismatic stacking fault with local sequence ABCD**AB**ABCD. The existence of this fault has been confirmed by first-principles calculations [12, 17, 18, 19]. Owing to its pure-edge character, this loop contains two possible straight edge segments, with line directions $\langle 2\bar{1}\bar{1}0\rangle$ and $[0001]$.

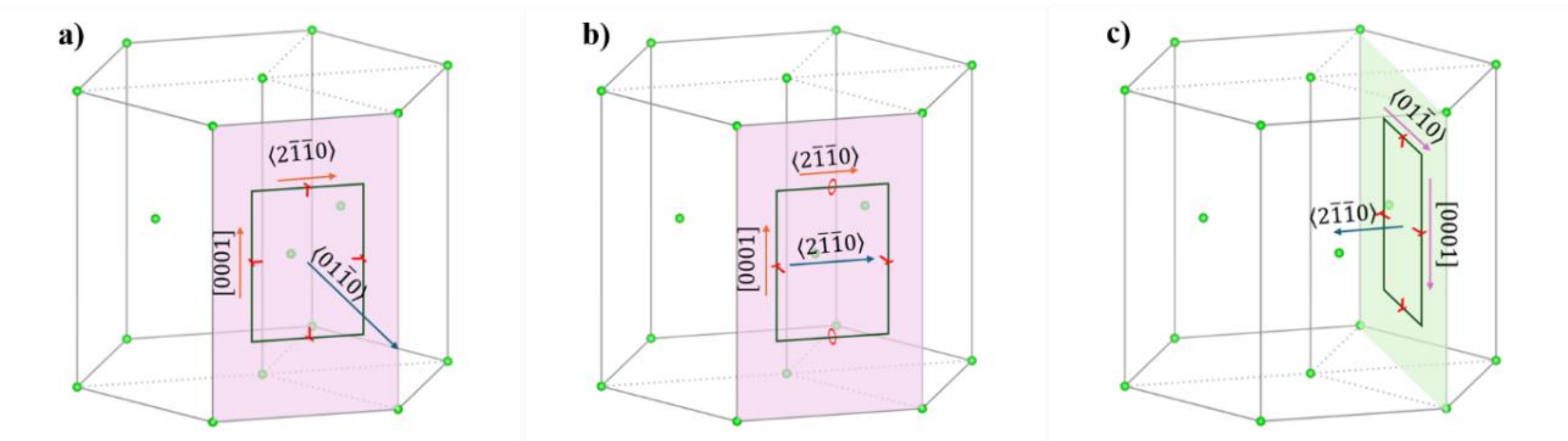

Figure 1: Dislocation characteristics of the three possible prismatic ⟨a⟩ loops, including their Burgers vectors,

habit planes and constituent line segments: (a) faulted first-order prismatic loop, (b) perfect first-order prismatic loop and (c) perfect second-order prismatic loop.

Because the embedded prismatic stacking fault is metastable, the faulted first-order prismatic ⟨a⟩ loop is expected to unfault by shear along $1/6\langle 2\bar{1}\bar{1}0\rangle$, transforming into a perfect first-order prismatic ⟨a⟩ loop (Fig. 1b) through the reaction [12, 17, 18, 19]:

$$\frac{1}{2}\langle 10\bar{1}0\rangle + \frac{1}{6}\langle \bar{1}2\bar{1}0\rangle \rightarrow \frac{1}{3}\langle 11\bar{2}0\rangle$$

After this transformation, the Burgers vector changes to $1/3\langle 11\bar{2}0\rangle$, which lies in the habit plane, giving the loop mixed edge–screw character and making it the glissile ⟨a⟩ loop in Zr. Its principal segments are a screw segment along $\langle 11\bar{2}0\rangle$ and an edge segment along $[0001]$. the change in Burgers vector also realigns the extra half-plane from a first-order to a second-order prismatic plane. The third possible subclass is the perfect second-order prismatic ⟨a⟩ loop (Fig. 1c), which is pure edge in character and has edge segments along [0001] and $\langle 10\bar{1}0\rangle$. Previous theoretical work suggests that this loop may form either directly from vacancy or interstitial cluster collapse, analogous to the faulted first-order loop, or through a rotation or transformation from first-order prismatic ⟨a⟩ loops, although the detailed transition mechanism remains unclear [12].

Although the three ⟨a⟩-loop subclasses described above have been proposed by atomistic simulations and compared through systematic energetic analyses, the experimental–simulation link remains selective rather than comprehensive. TEM studies have most commonly identified perfect first-order prismatic ⟨a⟩ loops with Burgers vector $1/3\langle 11\bar{2}0\rangle$, whereas direct experimental evidence for faulted first-order and perfect second-order prismatic ⟨a⟩ loops remains limited [9, 11, 15, 16]. This is partly because key loop characteristics, including vacancy or interstitial nature, mixed or pure-edge character, and the presence of a metastable prismatic stacking fault, are difficult to resolve experimentally [20, 21]. In particular, faulted first-order ⟨a⟩ loops may unfault as they grow to TEM-detectable sizes, leaving perfect first-order ⟨a⟩ loops in the typical 5–20 nm observation range, while perfect second-order ⟨a⟩ loops are expected to be comparatively rare [11, 16, 22]. In parallel, experimental studies have increasingly focused on radiation-induced segregation and solute clustering at ⟨a⟩ loops [9, 10, 23, 24, 25].

A complete atomistic description of dislocation loops in alloyed α-Zr requires both accurate short-range defect chemistry and a reliable long-range elastic field [12, 26, 27, 28]. First-principles density functional theory (DFT) provides the electronic-level accuracy needed to describe dislocation-core bonding, solute binding, and local chemical reconstruction, but the large simulation cells required to capture dislocation elastic fields, often exceeding thousands of atoms, make direct DFT calculations impractical [27, 28, 29, 30]. In contrast, empirical or semi-empirical molecular dynamics potentials such as embedded-atom-method (EAM) can access larger defect structures and their elastic fields, but their chemical fidelity is often insufficient for describing solute–core interactions and transition pathways [31, 32, 33]. To overcome this limitation, our previous work developed a hybrid quantum-mechanics/machine-learning forcefield (QM/ML) framework that combines DFT-level chemical accuracy with the efficiency of neural-network-based machine-learning interatomic potentials (MLIPs) [34, 35, 36]. This QM/ML framework enables high-accuracy dislocation studies beyond static energetic comparisons, including core-structure evolution, solute-induced reconstruction, and transition-barrier calculations [34]. Here, we extend the application of this framework to dislocation-core evolution, solute-induced rearrangement and QM/ML-NEB transition-energy profiles, including unfaulting and prismatic glide of first-order ⟨a⟩-loop segments.

In the present work, we apply this QM/ML framework to systematically investigate all nine possible line-segment configurations associated with the three ⟨a⟩-loop subclasses in both vacancy and interstitial forms. We evaluate the binding of representative alloying and impurity species, including interstitial Fe and H and substitutional Nb and Sn, to each dislocation-core geometry. Selected case studies further reveal how solute binding can modify local core structure, promote partial-dislocation interactions, or alter recombination behavior. Finally, nudged elastic band (NEB) calculations are performed to quantify the transition barrier for the faulted-to-perfect first-order ⟨a⟩-loop transformation and the Peierls barriers of perfect first-order ⟨a⟩ loop segments prismatic glide, providing an atomistic basis for assessing loop stability, mobility, and solute-modified mechanical response in irradiated α-Zr alloys.

## 2. Methods

### 2.1 Hybrid QM/ML setup

The full QM/ML workflow, including the pre-QM/ML benchmarking and code development, is described in our previous work [34]. Briefly, the present calculations combine DFT, performed using the Vienna Ab initio Simulation Package (VASP), with the pretrained MACE-MPA-0 medium model, in single-precision format [37, 38, 39]. This hybrid strategy is designed to retain DFT-level accuracy in the short-range dislocation-core and solute-interaction region, while using the MLIP to describe the surrounding elastic response of the Zr dislocation system efficiently. All QM/ML calculations were driven through the Python Atomic Simulation Environment (ASE) [40]. Detailed MLIP benchmark results are provided in the Supporting Information. One particularly relevant metric is the energy of the metastable prismatic stacking fault. DFT predicts a generalised stacking-fault energy of 190.8 mJ $m^{-2}$, whereas the MLIP predicts 261.4 mJ $m^{-2}$, corresponding to an overestimation of 70.7 mJ $m^{-2}$. The limited representation of this uncommon fault structure in typical training datasets may contribute to the difficulty of describing it accurately using MLIPs.

For the DFT region, the projector-augmented wave (PAW) method was used together with the Perdew–Burke–Ernzerhof (PBE) exchange–correlation functional [41, 42]. A plane-wave cutoff energy of 400 eV was employed. Methfessel–Paxton smearing with a width of 0.2 eV was used, together with a 1 or 6 k-point mesh scheme, where six k points were applied along the periodic dislocation-line direction and single k-point sampling was used in the two nonperiodic directions for each system. These settings were chosen to balance electronic convergence with the computational cost required for high-throughput dislocation–solute calculations [34]. The electronic convergence threshold was set to $1 \times 10^{-6}$ eV for each DFT step. Real-space projection of the PAW projectors was used for all calculations. Periodic boundary conditions were applied along the dislocation-line direction, while vacuum regions were introduced in the two directions normal to the dislocation line. The PAW valence configurations were Zr $4s^2 4p^6 4d^3 5s^1$, Nb $4p^6 4d^4 5s^1$, Sn $5s^2 5p^2$, Fe $3d^7 4s^1$, and H $1s^1$.

MLIP pre-relaxations were performed using the fast inertial relaxation engine (FIRE) algorithm [43]. A maximum force convergence criterion of 0.01 eV $Å^{-1}$ was used for MLIP-only pre-relaxations, while a criterion of 0.05 eV $Å^{-1}$ was used for the subsequent QM/ML production relaxations. For QM/ML-NEB calculations, both endpoint structures were first relaxed using the same QM/ML scheme. Intermediate NEB images were then generated between the relaxed endpoints and pre-relaxed using the MLIP to a maximum force criterion of 0.07 eV $Å^{-1}$, followed by QM/ML-NEB optimization with a target force criterion of 0.1 eV $Å^{-1}$. Owing to the high cost of QM/ML-NEB calculations, a looser convergence criterion is used here. These NEB results are therefore used primarily for qualitative comparison between closely related configurations.

For the solute-binding calculations, a random-sampling script was used to generate 15 non-repeating substitutional or interstitial candidate sites within a 5 Å radius of the dislocation core, all lying within the central quantum-mechanical region. The region partition and the energy and force assignments are summarised in Fig. S1. This procedure provides broad coverage of the chemically relevant local environments around the dislocation core while avoiding excessive redundancy from symmetry-equivalent sites. For each solute species and dislocation segment, the configurations with the lowest and highest MLIP predicted binding energies among the 15 sampled sites were selected for QM/ML relaxation and binding-energy evaluation. These provide contrasting sampled environments; they do not establish the global minimum and maximum of the QM/ML binding energy. More specifically, for each of the nine dislocation line segments considered in this work, solute–dislocation binding energies were calculated for interstitial Fe, interstitial H, substitutional Nb, and substitutional Sn. Including the solute-free reference calculations, this gives a total of 81 QM/ML calculations for the ⟨a⟩-loop line-segment–solute interaction study: nine solute-free dislocation calculations and 72 solute-containing calculations.

The group of binding-energy calculations involving the ⟨a⟩-type screw dislocation segment, indexed as as1 and presented in Figs. 3–7, was previously reported as a representative alloy-system application of the newly developed QM/ML code [34]. These results are included here for completeness because they were obtained using the same computational workflow and are required for a systematic comparison across all ⟨a⟩-loop line segments. Whereas Ref. [34] used the as1 calculations primarily to demonstrate the methodology, the present study incorporates them into the complete dataset and provides a broader comparative analysis and physical interpretation of the binding-energy behaviour across the full set of line-segment configurations.

## 2.2 Dislocation generation and visualisation

Initial dislocation configurations were generated using Atomsk [44]. All dislocation systems were constructed as cylindrical dislocation disks with a radius of 100 Å. Periodic boundary conditions were applied along the dislocation-line direction, and at least four atomic layers were included in this direction to reduce artificial solute–solute interactions across the periodic boundary. For dislocation disks with line directions along $\langle 11\bar{2}0\rangle$ or $\langle 10\bar{1}0\rangle$, atoms within a central radius of 10 Å from the dislocation core were assigned to the pure DFT region. This region was surrounded by an 8 Å QM/ML buffer region, in which DFT and MLIP contributions were smoothly coupled to provide a stable force distribution during QM/ML relaxation. For dislocation disks with line direction along $[0001]$, the pure DFT radius was reduced to 8 Å and the buffer thickness to 7 Å because of the higher atomic density associated with this orientation. These choices limited the DFT-treated region to approximately 250–300 atoms, providing a practical balance between computational cost and chemically accurate description of the dislocation core and nearby solute environments.

Two approaches were used to generate dislocation structures. The first approach applies the anisotropic linear-elastic displacement field to the initial crystal [44, 45]. This method preserves the total number of atoms and the atomic indexing, making it particularly suitable for generating screw dislocations and for constructing Peierls-barrier endpoint structures. Because NEB calculations require consistent atom identities and ordering across all images, the linear-elastic displacement method provides a convenient way to create initial and final states corresponding to the same dislocation translated between adjacent lattice positions. However, for edge dislocations in finite disk geometries, this method can introduce a surface step at the outer boundary of the simulation cell, leading to artificial surface contributions to the total energy. In addition, because it only displaces atoms without inserting or removing lattice planes, it cannot directly generate stacking-fault-containing dislocation configurations.

The second approach generates edge dislocations by inserting or removing a half-plane of atoms. This construction is more directly connected to irradiation-induced vacancy or interstitial loop formation and was therefore used for edge-dislocation solute-binding calculations and for stacking-fault unfaulting transition studies. In particular, faulted first-order prismatic ⟨a⟩-loop segments were generated using the half-plane construction followed by structural relaxation. During preliminary MLIP-only relaxation, stacking-fault-containing configurations showed a tendency to unfault spontaneously, which is consistent with the sensitivity of these structures to the predicted prismatic stacking-fault energy. This behavior further motivates the use of the QM/ML framework, in which the core-region transition is governed by DFT-level energetics rather than by the MLIP alone.

Structural analysis and visualization were performed using OVITO [46]. The dislocation extraction algorithm (DXA) was used to monitor dislocation character, line direction, and Burgers vector where applicable [47, 48]. However, DXA has limitations for the present α-Zr systems because prismatic stacking faults and faulted ⟨a⟩ loops with Burgers vector $1/2\langle 10\bar{1}0\rangle$ are not always recognized as conventional dislocation lines. In these cases, the faulted loop may instead appear as a thin defect mesh or planar defect region. Therefore, DXA analysis was supplemented by direct inspection of atomic configurations, stacking-sequence changes, and local core structures when identifying faulted first-order prismatic ⟨a⟩ segments.

## 2.3 Defect binding energy and NEB energy profile evaluation

The solute–dislocation binding energy in the QM/ML framework is evaluated as,

$$E_{QM/ML}^{binding} = (E_{QM/ML}^{dislo+sol} - E_{QM/ML}^{dislo}) - (E_{DFT}^{Zr+sol} - E_{DFT}^{Zr}) \quad (3)$$

Two QM/ML simulations are required: a dislocation disk containing the solute ($E_{QM/ML}^{dislo+sol}$) and the corresponding solute-free reference ($E_{QM/ML}^{dislo}$). The DFT energy difference for a solute in a perfect hcp Zr cell is subtracted to isolate the interaction energy between the dislocation core and the solute.

For QM/ML-NEB simulations, five intermediate images were inserted between the relaxed endpoint configurations. For Peierls barrier calculations, the endpoints correspond to the same dislocation disk with matching atoms but with the dislocation core located in adjacent Peierls valleys. The NEB coordinate λ was written as:

$$\lambda = \frac{\xi}{a_0}$$

where $\xi$ is the physical dislocation displacement along the glide direction, $a_0$ is the distance between adjacent equivalent Peierls valleys along the glide direction. Thus, $\lambda = 0$ and $\lambda = 1$ refer to the initial and final core positions. Notably, as discussed previously for the construction of the dislocation disks, the high-energy surface atom geometry can differ slightly between the two endpoints, introducing artificial energy contributions. This effect is particularly important for the Peierls barrier evaluation. Thus, a linear endpoint-energy contribution was removed from the NEB energy profile. The corresponding Peierls energy of images $i$ was then:

$$\Delta E_i^P = \frac{E_i - [(1-\lambda_i)E_0 + \lambda_i E_1]}{L}$$

where $E_i$ is the energy of image $i$, $E_0$ and $E_1$ are the endpoint energies, and $L$ is the dislocation line length along the disk axis. The Peierls barrier per unit length is then:

$$\Delta E_P = {}^{max}_{i}\Delta E_i^P$$

In this work, the Peierls energy profiles are reported in meV Å$^{-1}$ as a function of $\xi/a_0$.

For transitions of dislocations containing a prismatic stacking fault, energy profiles were calculated using the same NEB procedure, but $\lambda \in [0,1]$ was treated as a normalised reaction coordinate rather than as $\xi/a_0$. No surface-energy correction was applied because part of the surface-energy change is associated with the geometric transition of the dislocation itself. This is also why these results are treated mainly as a qualitative comparison. The relative energy was normalised by the stacking-fault area $A_{SF}$, defined as the disk radius multiplied by its length along the dislocation line. The energy is then expressed as:

$$\Delta\gamma_i = \frac{E_i - E_0}{A_{SF}}$$

The stacking-fault transition energy profiles are therefore reported in meV Å$^{-2}$.

## 2.4 Summary of dislocation line segments considered

Table 1 briefly summarises all dislocation line segments considered in this work. These include four edge dislocations containing prismatic stacking faults, with indices beginning with ae1; two edge components shared by the first and second perfect ⟨a⟩ loops; two edge components unique to the second perfect ⟨a⟩ loop; and the only screw dislocation in the first perfect ⟨a⟩ loop. Both interstitial- and vacancy-type edge dislocations are considered. Edge components within the same loop are distinguished by their line directions while retaining the same Burgers vector.

Table 1: Labelling indices and dislocation characters of the nine dislocation line segments considered across the three possible ⟨a⟩ loops.

| Index | Vacancy or interstitial | Screw or edge | Loop type | Burgers vector | Line direction |
|---|---|---|---|---|---|
| ae1i0 | Interstitial | Edge | 1st faulted | $1/2\langle 10\bar{1}0\rangle$ | $[0001]$ |
| ae1v0 | Vacancy | Edge | 1st faulted | $1/2\langle 10\bar{1}0\rangle$ | $[0001]$ |
| ae1i1 | Interstitial | Edge | 1st faulted | $1/2\langle 10\bar{1}0\rangle$ | $\langle 11\bar{2}0\rangle$ |
| ae1v1 | Vacancy | Edge | 1st faulted | $1/2\langle 10\bar{1}0\rangle$ | $\langle 11\bar{2}0\rangle$ |
| ae2i0 | Interstitial | Edge | 1st and 2nd perfect | $1/3\langle 11\bar{2}0\rangle$ | $[0001]$ |
| ae2v0 | Vacancy | Edge | 1st and 2nd perfect | $1/3\langle 11\bar{2}0\rangle$ | $[0001]$ |
| ae2i-1 | Interstitial | Edge | 2nd perfect | $1/3\langle 11\bar{2}0\rangle$ | $\langle 10\bar{1}0\rangle$ |
| ae2v-1 | Vacancy | Edge | 2nd perfect | $1/3\langle 11\bar{2}0\rangle$ | $\langle 10\bar{1}0\rangle$ |
| as1 | ------------- | Screw | 1st perfect | $1/3\langle 11\bar{2}0\rangle$ | $\langle 11\bar{2}0\rangle$ |

# 3. Results

## 3.1 Dislocation line geometry and partial dislocation dissociation

Before examining defect energetics, the microstructures of all nine line segments after QM/ML relaxation provide a basis for understanding the evolution of core geometry in loops, particularly the dissociation of dislocations into partials, as summarised in Table 2. For the faulted loops, labelled with indices beginning with 'ae1', no clear partial separation is identified for dislocations with line directions along $[0001]$. However, the core geometry and defect mesh obtained from DXA indicate a slight tendency for the dislocation core to extend onto another first-order prismatic plane oriented at 120° to the faulted plane. As OVITO does not fully identify the faulted dislocation, the corresponding partial-separation behaviour cannot be determined directly. Further insight into this core geometry is obtained from the strain field introduced by solute atoms, as discussed in the following section.

Table 2: Summary of the dissociation behaviour of all dislocation line segments following QM/ML relaxation.

| Index | Dissociation or not | Perfect Burgers vector | Partial 1 Burgers vector | Partial 2 Burgers vector |
|---|---|---|---|---|
| ae1i0 | No | $1/2[1\bar{1}00]$ | -------------- | -------------- |
| ae1v0 | No | $1/2[1\bar{1}00]$ | -------------- | -------------- |
| ae1i1 | Yes | $1/2[1\bar{1}00]$ | $1/3[1\bar{1}00]$ | $1/6[1\bar{1}00]$ |
| ae1v1 | Yes | $1/2[1\bar{1}00]$ | $1/3[1\bar{1}00]$ | $1/6[1\bar{1}00]$ |
| ae2i0 | Yes | $1/3[1\bar{2}10]$ | $1/3[1\bar{1}00]$ | $1/3[0\bar{1}10]$ |
| ae2v0 | Yes | $1/3[1\bar{2}10]$ | $1/3[1\bar{1}00]$ | $1/3[0\bar{1}10]$ |
| ae2i-1 | Yes | $1/3[1\bar{2}10]$ | $1/3[1\bar{1}00]$ | $1/3[0\bar{1}10]$ |
| ae2v-1 | Yes | $1/3[1\bar{2}10]$ | $1/3[1\bar{1}00]$ | $1/3[0\bar{1}10]$ |
| as1 | No | $1/3[1\bar{2}10]$ | -------------- | -------------- |

For the remaining seven line segments, the dislocation characters are more clearly resolved. For further detail, the atomistic configurations and DXA-resolved dislocation lines are presented for four representative systems that collectively cover the principal characteristics observed in this study. Fig. 2(a1) and (a2) show the sole screw-dislocation segment in the first perfect loop. DXA identifies the full Burgers vector, with no evident dissociation into partial dislocations. A previous first-principles DFT and EAM study of this screw component proposed dissociation into two partials belonging to the $1/6\langle 11\bar{2}0\rangle$ Burgers-vector family, although this dissociation is not captured by the DXA-resolved line character. Nevertheless, the elongated dislocation-core region, indicated by the central white atoms with fault-like character, is consistent with the core geometry proposed in that study. DXA does not resolve separate partials in this core, so the present analysis does not establish their separation. The discussion therefore focuses on partial separations supported by the atomic configurations and dislocation analysis.

The ae1i0 and ae1v0 dislocations show no clear partial dissociation and are not fully resolved as distinct dislocation lines. The other two faulted dislocations, ae1i1 and ae1v1, provide further insight into their core structures, as shown in Fig. 2(b1) and (b2). In addition to an unidentified partial represented by the defect mesh, DXA identifies a dissociated Shockley partial with Burgers vector $1/3[1\bar{1}00]$. As the two partial Burgers vectors sum to the full Burgers vector $1/2[1\bar{1}00]$, a possible dissociation reaction can be written as:

$$\frac{1}{2}[1\bar{1}00] \rightarrow \frac{1}{3}[1\bar{1}00] + \frac{1}{6}[1\bar{1}00]$$

The $1/6[1\bar{1}00]$ partial dislocation is identified with the aid of the dislocation microstructural evolution observed during the subsequent stacking-fault transition attempts. A more detailed discussion is presented in the corresponding section. Returning to the dissociation behaviour identified here, the stabilisation mechanism of the metastable prismatic dislocations containing a stacking fault becomes clearer. A second intrinsic basal stacking-fault ribbon (SF ribbon) is identified between the partials along the $[0001]$ direction. A similar stacking-fault ribbon between Shockley partials was reported in a previous DFT study, while the associated edge-dislocation configuration is supported here.

The remaining two groups of subplots show the edge components of the perfect loops. Fig. 2(c1) and (c2) present the core geometry and dislocation character of the ae2i-1 and ae2v-1 segments. As for the ae1i1 and ae1v1 segments, the perfect dislocation dissociates into partials. In this case, two Shockley partials with different directions are identified, together with a locally recombined perfect-dislocation segment in the enlarged view in Fig. 2(c2). The two resulting Shockley partials are inclined at 60° to one another and follow the dissociation reaction:

$$\frac{1}{3}[1\bar{2}10] \rightarrow \frac{1}{3}[1\bar{1}00] + \frac{1}{3}[0\bar{1}10]$$

A similar SF ribbon with a second intrinsic stacking-fault structure is identified along $[0001]$.

The final dislocation group comprises the edge components shared by the first and second perfect ⟨a⟩ loops, with line directions along $[0001]$, denoted ae2i0 and ae2v0 (Fig. 2(d1) and (d2)). In these cases, DXA identifies only a perfect dislocation and does not resolve the Shockley-partial dissociation observed for the other perfect-loop segments. This is more likely to reflect a limitation of DXA than the absence of dissociation. Examination of the atomistic core geometry reveals a distinct stacking-fault region elongated along $\langle 11\bar{2}0\rangle$, with the fault extending along $\langle 1\bar{1}00\rangle$. In the previously identified Shockley-dissociation cases, the basal SF ribbons extend along $[0001]$. Here, because the dislocation line is itself parallel to $[0001]$, the corresponding ribbon instead has prismatic stacking-fault character, which, as discussed above, is not fully recognised by DXA.

Closer inspection of the core geometry provides further evidence of dissociation. Formation of the apparent perfect dislocation requires the addition or removal of two atomic planes. After rotating the viewing direction by 30°, a partial dislocation associated with one terminating atomic half-plane can be identified, together with a second partial viewed after a 30° rotation in the opposite direction. These features are consistent with the Shockley-partial configurations identified for the other perfect dislocations. The atomistic geometry therefore supports consistent dissociation behaviour across the perfect dislocations, despite the limited resolution of DXA for the prismatic stacking-fault structure.

Overall, the dissociation reactions and associated core geometries are identified and visualised. The results are consistent with previously proposed configurations while extending the analysis to different dislocation-line directions and characters. The interstitial or vacancy nature of the edge dislocations produces no evident difference in the relaxed core geometry or dissociation reaction. However, differences emerge under solute-induced strain fields and during manually promoted transitions, as discussed in the following sections.

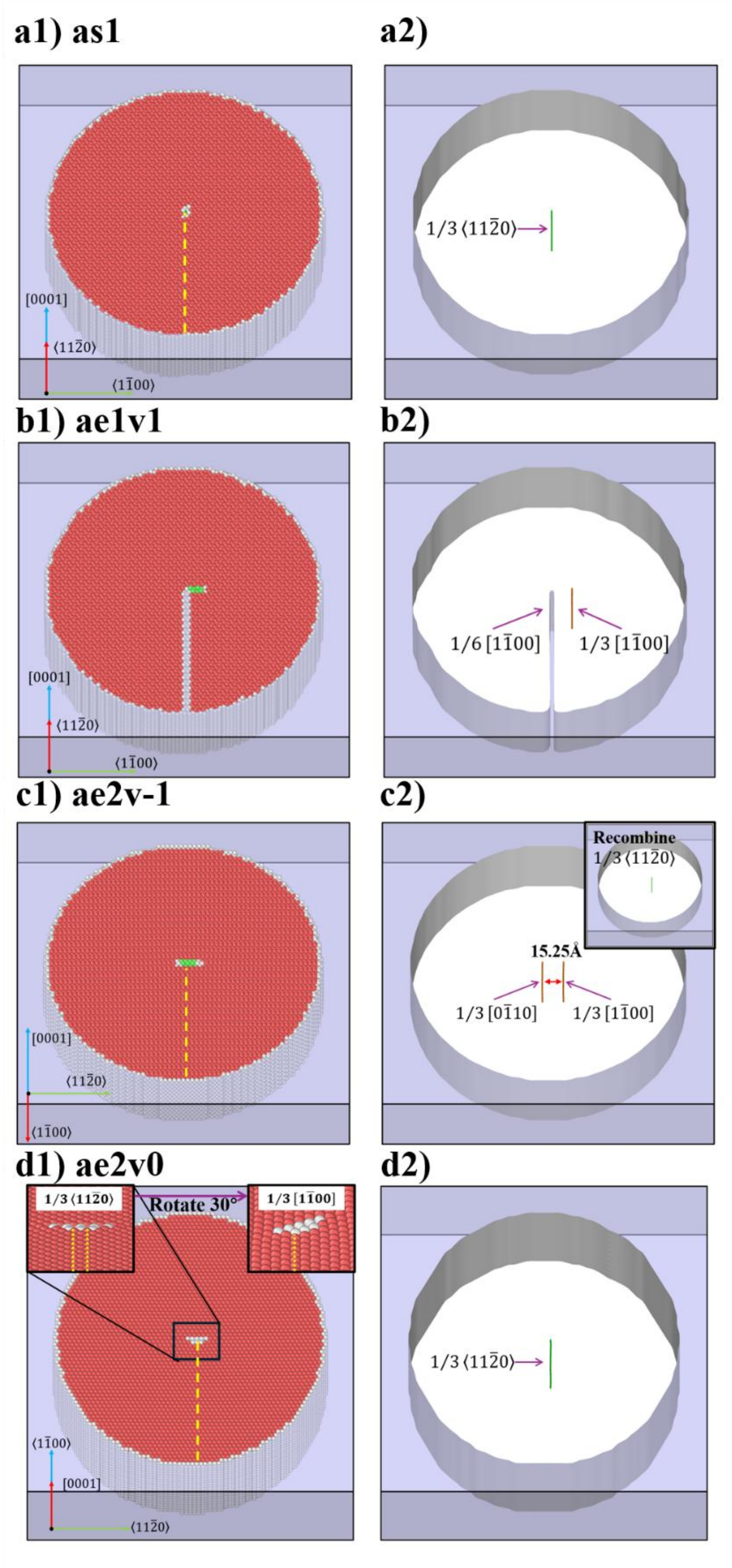

Figure 2: (1) Atomic configurations of the dislocation discs, with red atoms representing the hcp lattice and white and green atoms at the core indicating possible prismatic- or basal-fault crystallography. Yellow dashed lines mark the associated atomic half-planes of the edge dislocations. (2) Corresponding DXA-resolved dislocation lines and Burgers-vector labels for each configuration in the same column. The dislocations shown are (a) as1, (b) ae1v1, (c) ae2v-1 and (d) ae2v0.

## 3.2 Solute-dislocation binding energies

Of the 81 QM/ML geometry-optimisation calculations performed, nine correspond to reference states, leaving 72 solute–dislocation binding-energy calculations. Before considering individual solute species, the general preferences for different dislocation-loop line segments and for interstitial- and vacancy-type segments are compared in Fig. 3. The QM/ML box plots show binding energies that include both DFT and MLIP contributions and use the corresponding DFT reference states. The DFT box plots show interaction energies evaluated for the finite DFT cluster within the QM/ML-relaxed configuration. For comparison, the MLIP box plots show energies obtained from structures pre-relaxed entirely using the MLIP, with the corresponding MLIP reference states and no subsequent QM/ML relaxation. The finite-cluster DFT value provides a reference for the local chemically sensitive interaction, whereas the QM/ML value also includes the surrounding MLIP contribution. Their difference can therefore contain both longer-range interactions and residual embedding error; it is not, by itself, a measure of MLIP error. The MLIP-only comparison additionally includes differences in the relaxed geometry [34].

Overall, the binding energies associated with faulted-loop dislocations show a wider spread across all three methods. A larger proportion of these values are also more negative, indicating stronger attractive interactions than those obtained for the perfect-loop segments. One notable exception is an interaction energy of approximately −3 eV for interstitial Fe binding to the ae2i0 edge segment shared by the perfect loops. This strongly attractive interaction is reproduced across the three energy definitions and is examined in detail in the following section.

More generally, the mean binding energies appear slightly more negative for interstitial-type dislocations than for vacancy-type dislocations. However, some vacancy-type segments exhibit strongly repulsive interactions, with positive binding energies approaching +1 eV for particular solutes. These extreme values are assessed separately for each solute species in the following analysis.

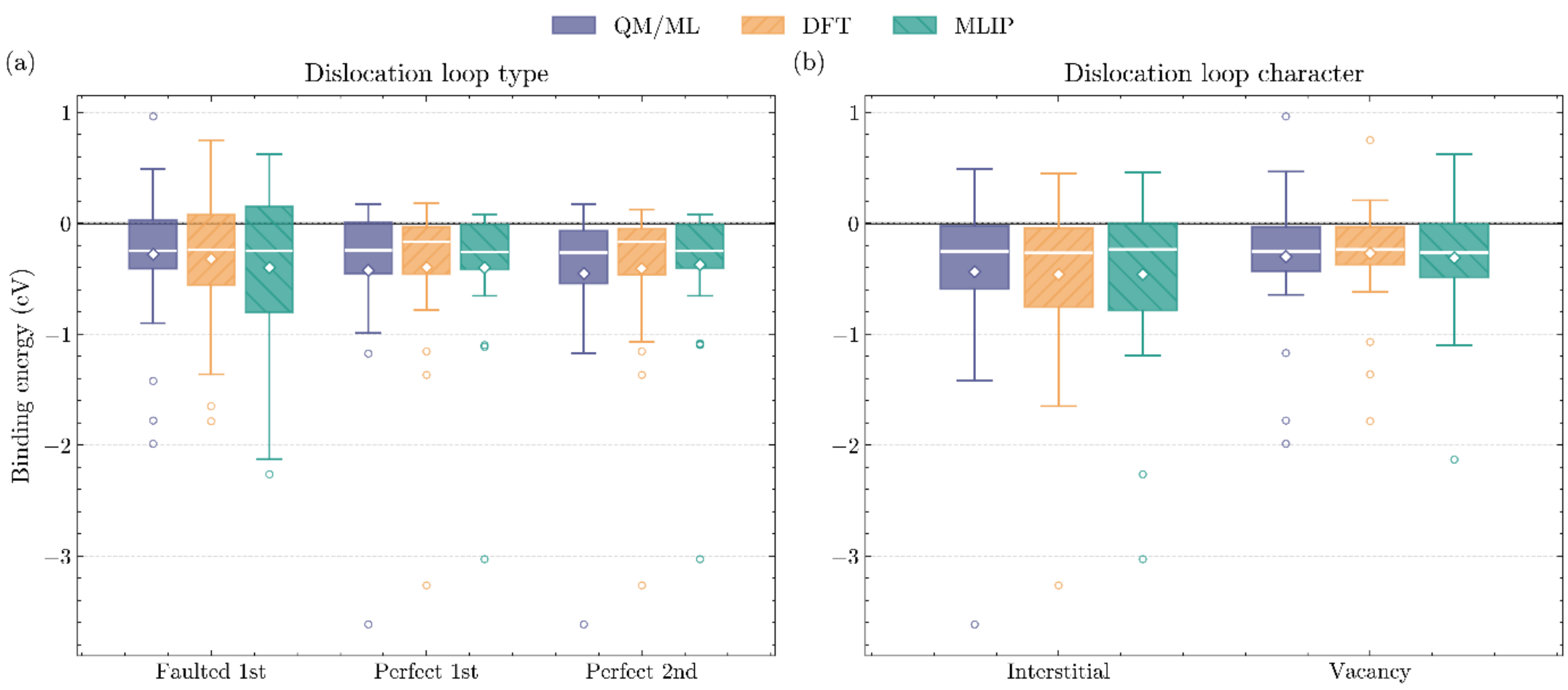

Figure 3: Solute-dislocation binding energies for all solute species combined, grouped according to (a) the loop type associated with each dislocation line segment and (b) the interstitial/vacancy type character of the corresponding dislocation. Whiskers were defined using the conventional 1.5×IQR criterion. The as1 binding-energy results were previously reported in Ref. [34] and are included here for systematic comparison.

Fig. 4 presents the solute-resolved binding preferences for the three loop types. Overall, the interstitial solutes Fe and H exhibit predominantly negative binding energies around the dislocation cores, indicating attractive interactions. Fe shows particularly strong attraction, consistent with experimentally observed radiation-induced segregation. In contrast, the substitutional solutes Sn and Nb can exhibit strongly repulsive interactions, particularly with the faulted-loop segments.

Excluding the few extreme values and considering the median and mean of each distribution, Fe and H show a stronger preference for the faulted loop than for the perfect loops formed after the faulted-loop transition. However, the upper, less attractive range of the distributions remains broadly comparable among the loop types. By contrast, Sn and Nb preferentially bind to the perfect loops rather than to the faulted loop, with this trend being particularly pronounced for Sn. Although Nb can be repelled by the faulted loop, strong repulsion is not generally observed for the perfect loops, as indicated by the upper binding-energy range remaining close to 0 eV. Sn, however, can still exhibit repulsive interactions with the perfect loops.

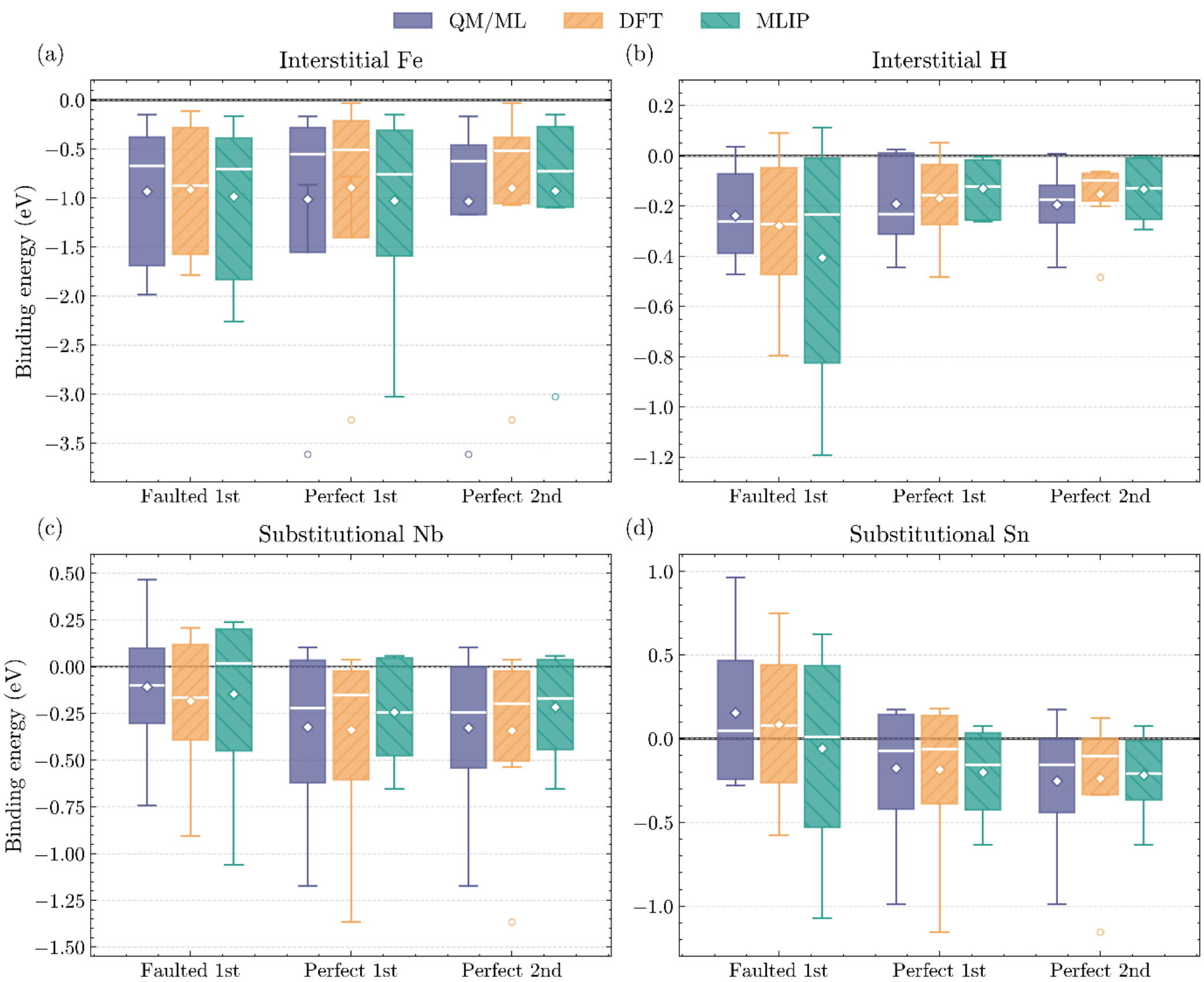

Figure 4: Solute-dislocation binding energies grouped according to the loop type associated with each dislocation line segment for (a) interstitial Fe, (b) interstitial H, (c) substitutional Nb and (d) substitutional Sn. Whiskers were defined using the conventional 1.5×IQR criterion. In subplot (a), all observations, including the extreme value, were retained; the resulting extended plotting range visually compresses the central distributions of two box-and-whisker plots but does not affect the underlying numerical values. The as1 binding-energy results were previously reported in Ref. [34] and are included here for systematic comparison.

In addition to the loop type associated with each line segment, the three possible line directions, [0001], $\langle 11\bar{2}0\rangle$ and $\langle 10\bar{1}0\rangle$, provide another important basis for comparison, as shown in Fig. 5. As discussed in the previous section, the line direction influences the dislocation-core geometry even within the same loop type. For example, perfect edge dislocations with the same Burgers vector and the same dissociation reaction into two Shockley partials can form different SF ribbons depending on their line direction. The corresponding stacking-fault energies may therefore influence solute trapping. The binding preferences for the three line directions are compared here across all four solute species.

Overall, the $[0001]$ segments exhibit the widest spread in binding energies. With the inclusion of the DFT contribution, Fe can bind particularly strongly to these segments, although strong attraction is also observed for some of the other solutes. However, the $[0001]$ direction also produces some of the strongest repulsive interactions, particularly for Sn and Nb. The binding-energy distributions are considerably narrower for the $\langle 11\bar{2}0\rangle$ and $\langle 10\bar{1}0\rangle$ directions. 2/3 of the $\langle 11\bar{2}0\rangle$ segments are faulted edge dislocations, with the remaining segment having screw character, whereas all $\langle 10\bar{1}0\rangle$ segments belong to perfect loops. In general, the perfect $\langle 10\bar{1}0\rangle$ segments show more favourable and more consistent binding energies across the four solutes, although these segments occur only in the second perfect loop. By contrast, the predominantly faulted $\langle 11\bar{2}0\rangle$ segments show strong attraction to interstitial Fe, weaker but still generally attractive interactions with interstitial H, moderate attraction with limited repulsion for substitutional Nb, and limited attraction with more pronounced repulsion for Sn.

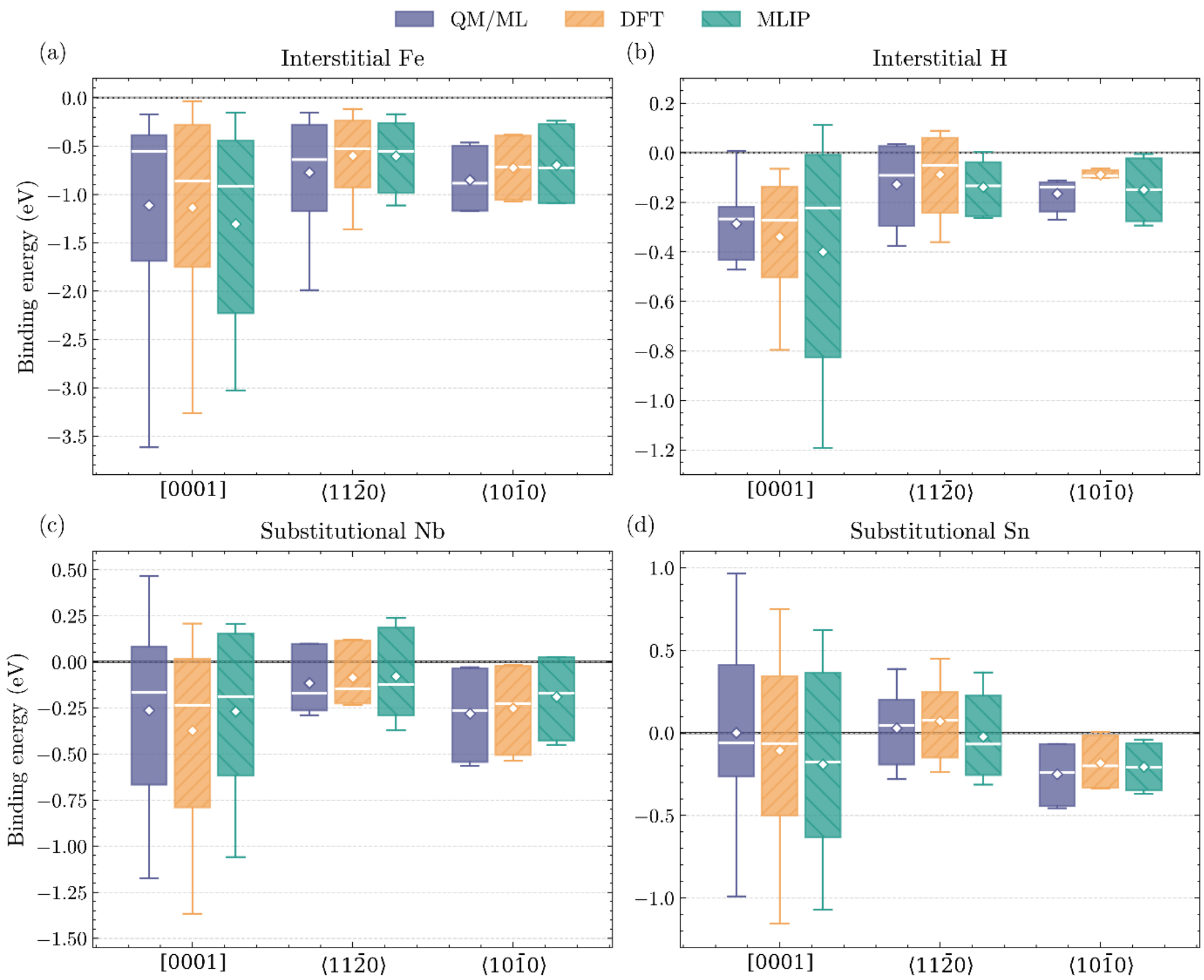

Figure 5: Solute-dislocation binding energies grouped according to dislocation-line direction for (a) interstitial Fe, (b) interstitial H, (c) substitutional Nb and (d) substitutional Sn. Whiskers were defined using the conventional 1.5×IQR criterion. The as1 binding-energy results were previously reported in Ref. [34] and are included here for systematic comparison.

The final comparison considers solute preferences for interstitial- and vacancy-type dislocations, as shown in Fig. 6. Notably, the as1 screw dislocation has no interstitial nor vacancy character, so the subset of energies are excluded here. Overall, the solutes show a stronger preference for interstitial-type dislocations, which produce more strongly attractive binding energies in most cases. Considering the upper limits of the distributions, which represent the least favourable interactions sampled, the interstitial solutes exhibit similar values for the two dislocation types. The substitutional solutes show more distinct behaviour. Nb is more strongly repelled by some vacancy-type than interstitial-type dislocations. For Sn, the largest positive outlier also occurs for a vacancy-type dislocation, although the central distributions differ from the extreme values.

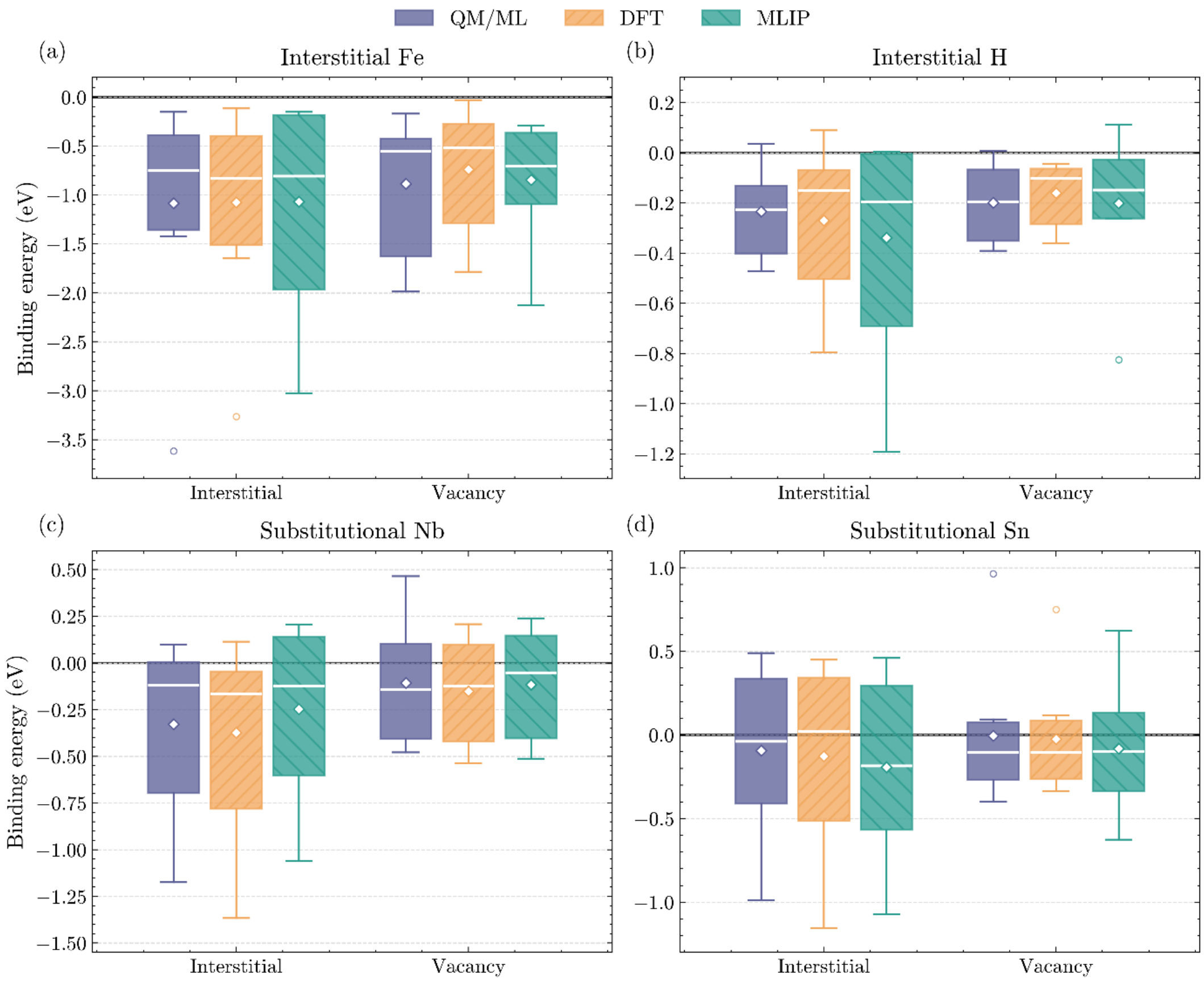

Figure 6: Solute-dislocation binding energies grouped according to the interstitial/vacancy type character of the dislocation for (a) interstitial Fe, (b) interstitial H, (c) substitutional Nb and (d) substitutional Sn. Whiskers were defined using the conventional 1.5×IQR criterion.

As noted above, the three approaches used to evaluate the binding energies do not always give consistent results for an individual set of calculations. Taking the DFT binding energy as the reference, because the dominant local interaction is evaluated within the DFT region, the deviations of the QM/ML and MLIP binding energies from DFT are compared in Fig. 7. A negative deviation indicates that the tested method predicts a more attractive binding energy than DFT, whereas a positive deviation indicates a less attractive or more repulsive interaction. These deviations may reflect errors in the MLIP arising from incomplete representation of the relevant physics or defect chemistry in the training set. They may also capture long-range solute–strain interactions outside the DFT region and therefore absent from the DFT-only binding energy. The rationale and validation of this analysis have been discussed in detail previously [34].

Among the four solutes, interstitial Fe shows the widest deviations in both the positive and negative directions. Some values exceed the approximate ±0.2 eV range considered plausible for interactions omitted from the finite DFT region, suggesting that the MLIP may not fully describe Fe–dislocation interactions. The MLIP deviations are also more broadly distributed than the corresponding QM/ML deviations, indicating that the DFT region partly suppresses spurious MLIP contributions. For H, Nb and Sn, the deviations after QM/ML relaxation are generally smaller, suggesting that part of the remaining difference may arise from long-range interactions. The MLIP-only results are more dispersed. For H, the MLIP-only calculations can overpredict both attraction and repulsion. Possible contributors include the representation of local strain, H–Zr bonding and charge redistribution. The present energy comparison does not separate these effects. The MLIP tends to overpredict repulsion for Nb, whereas repulsion for Sn can be underestimated.

The deviations also depend on the loop type. The QM/ML results for the perfect loops are generally closer to the DFT values, whereas larger deviations occur for the faulted loops. The MLIP-only results show the same trend, with greater deviations for faulted than for perfect loops. This is consistent with the current MACE model overpredicting the energy of the metastable prismatic stacking fault, as identified in the MLIP benchmark. This limitation is likely associated with the rarity or insufficient representation of this fault structure in the training data. The stacking-fault error may affect not only solute trapping but also the predicted unfaulting behaviour, because an artificially elevated stacking-fault energy can reduce the apparent transition barrier. This effect is examined further using QM/ML-NEB simulations.

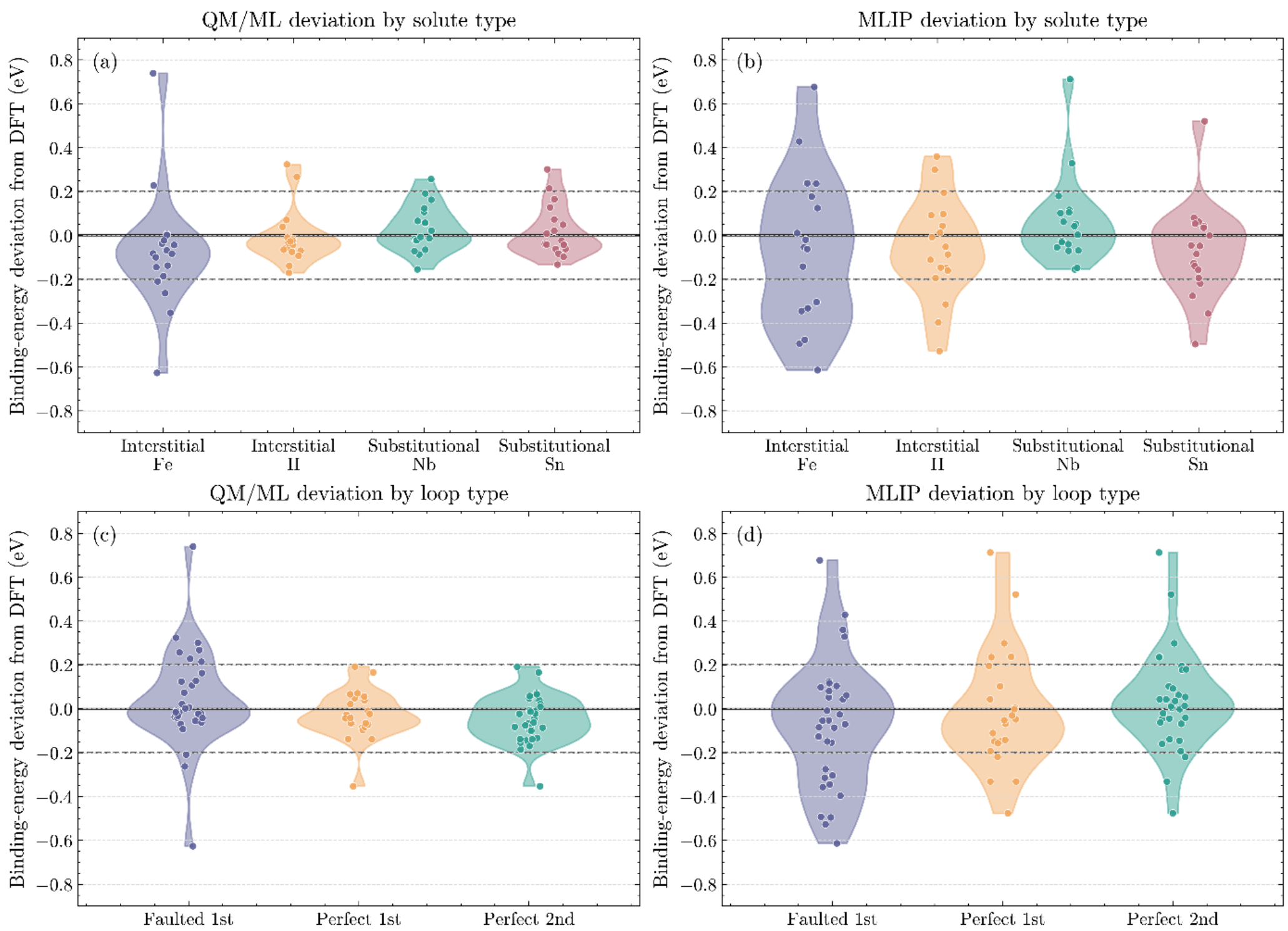

Figure 7: Deviations of the solute-dislocation binding energies from the DFT-only values for (a, c) QM/ML and (b, d) MLIP. The results are grouped by solute species in (a, b) and by the loop type associated with each dislocation line segment in (c, d). A more negative deviation indicates that the corresponding method predicts a more negative, and therefore more attractive, binding energy than the DFT-only calculation. The as1 binding-energy results were previously reported in Ref. [34] and are included here for systematic comparison.

### 3.3 Solute-induced changes in dislocation-core geometry

The solute–dislocation binding energies presented above provide a broad picture of how different solutes interact with ⟨a⟩ loop line segments, while also enabling quantitative comparison and cross-validation among the simulation methods. However, the energetic analysis alone does not describe how solute chemistry and the associated strain field influence the dislocation-core geometry. This section therefore examines notable features in representative solute–dislocation structures after QM/ML relaxation, supported by DXA analysis.

With reference to the solute-resolved binding energies in Figs. 4, 5 and 6, one particularly notable case is interstitial Fe binding to the ae2i0 segment, an interstitial-type edge segment with a line direction along $[0001]$. This configuration has a strongly attractive binding energy of −3.6 eV. The Fe solute substantially modifies the dislocation-core geometry, resulting in a lower-energy configuration. Changes in the position of the DXA-resolved line provide one measure of the associated core rearrangement. The DXA analysis used here provides the vertex coordinates of the resolved dislocation line, allowing the changes in these coordinates along the two radial directions of the simulation disc to be measured between the solute-free and solute-containing QM/ML-relaxed configurations. Table 3 presents representative configurations with large and small solute-induced core displacements, together with the corresponding QM/ML binding energies.

For this segment, larger displacement of the DXA-resolved dislocation line, indicating a greater change in core geometry, is associated with a more negative binding energy. The same general trend is observed not only for Fe, but also for H, which produces a comparatively small strain field, Nb, whose atomic radius is closest to that of Zr, and Sn. For each solute, greater core displacement corresponds to stronger attraction to the dislocation. In addition to the −3.6 eV Fe interaction, the lowest-energy H, Nb and Sn configurations for this segment are also the most attractive configurations found for these solutes across all ⟨a⟩ loop segments considered. These results highlight the strong solute affinity of this interstitial-type perfect-loop segment, which is associated with one of the experimentally observed conventional ⟨a⟩ loop configurations.

Table 3: Displacement of the DXA-resolved dislocation line and QM/ML binding energies for the two selected solute configurations in ae2i0.

| Solute | High E site drift Å | High E site binding eV | Low E site drift Å | Low E site binding eV |
|---|---|---|---|---|
| Fe-int | 1.71 | -0.45 | 2.80 | -3.62 |
| H-int | 0.53 | -0.21 | 1.95 | -0.44 |
| Nb-sub | 1.86 | 0.01 | 2.92 | -1.17 |
| Sn-sub | 0.01 | 0.17 | 3.18 | -0.99 |

Solute stabilisation associated with changes in dislocation-core geometry is also observed for several other line segments. For example, for the ae1v1 faulted edge segment, one Fe configuration reduces the width of the basal SF ribbon between the partials and displaces the Shockley partial towards the faulted partial by 5.5 Å. This configuration also exhibits a relatively strong binding energy of −1.1 eV. However, substantial displacement of partials bounding basal SF ribbons remains uncommon. All other solutes tested for ae1v1 and ae1i1 produce displacements below 1 Å. The present discussion of changes in partial-dislocation position in these less commonly studied faulted segments is therefore limited to this representative case. By contrast, the two faulted edge segments with line directions along $[0001]$, which are difficult to characterise using DXA alone, show more pronounced solute-induced changes.

Fig. 8 compares the configurations before and after QM/ML relaxation following the addition of a high-energy Fe solute site to the ae1i0 segment. As discussed above, the dislocation associated with the prismatic fault is not fully recognised by DXA and is therefore represented primarily by a defect mesh in Fig. 8(a1) and (a2). Nevertheless, the mesh indicates that the dislocation core is inclined towards the adjoining first-order prismatic plane at the centre of the simulation box. After addition of the solute, this inclined core appears to detach from the faulted half-plane. The central red atom between the dislocation core and the white defect-mesh region indicates that the local hcp structure has been restored in this region, as shown in Fig. 8(b1).

DXA then resolves a perfect dislocation separated from the remaining faulted dislocation. Faulted dislocations are expected to transform into perfect dislocations through shear on the prismatic plane. As no external shear is applied here, the result suggests that the combined effects of solute chemistry and local strain can promote separation of a perfect-loop component from the faulted configuration. This behaviour is observed not only for Fe, which produces a relatively large local strain field, but also for both tested H sites and for one low-energy Sn site. The occurrence across several solute configurations suggests that this is not an isolated interaction. However, the solute-induced transformation is observed only for the interstitial-type dislocation and not for the corresponding vacancy-type segment.

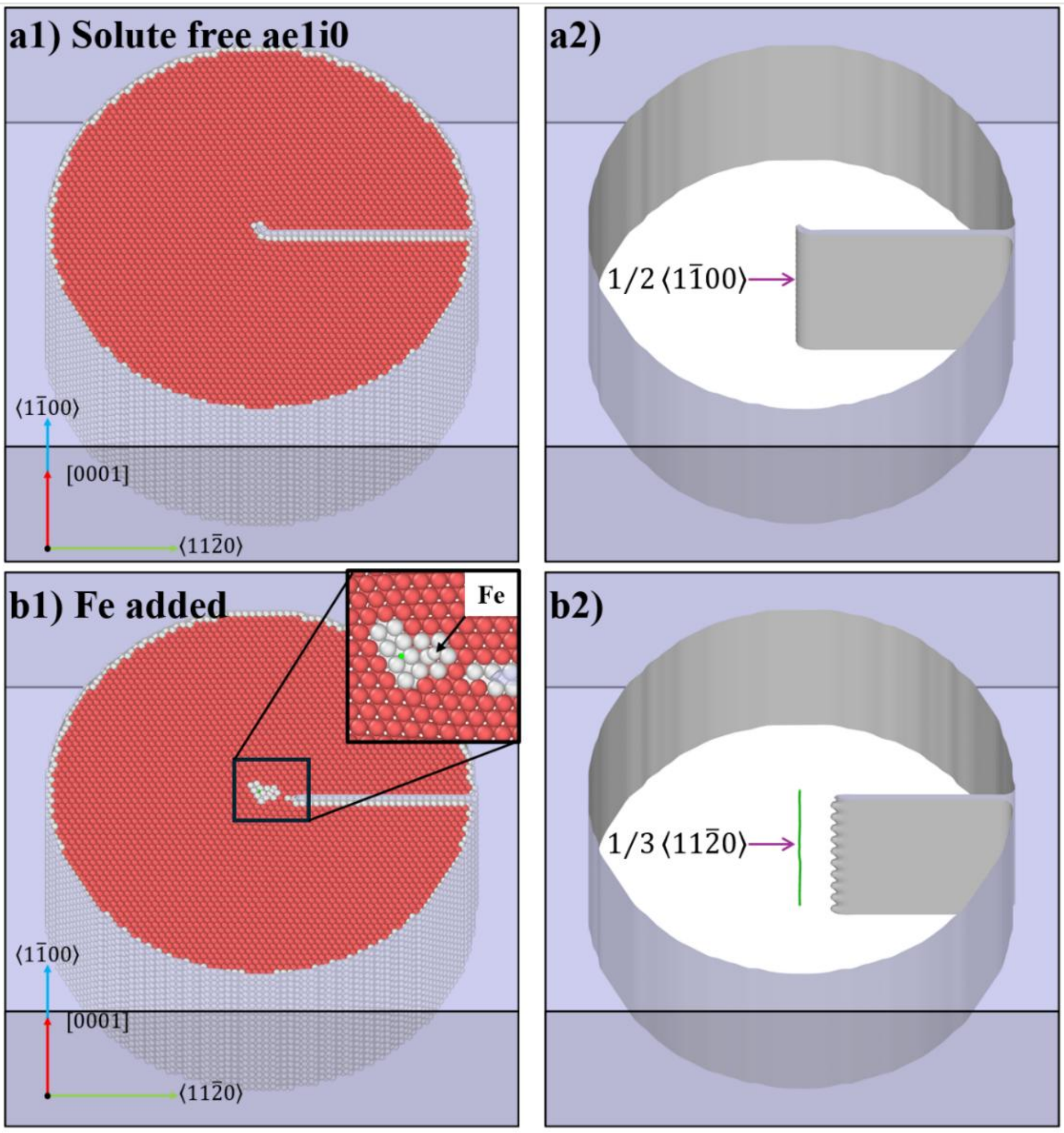

Figure 8: Case study of interstitial Fe interacting with the ae1i0 dislocation. (1) Atomic configurations of the dislocation discs and (2) corresponding DXA-resolved dislocation-line characters for (a) the solute-free dislocation and (b) the configuration containing interstitial Fe.

A final case study concerns ae2i-1, an edge segment of the second perfect loop. In Fig. 2(c1) and (c2), the corresponding vacancy-type segment, ae2v-1, clearly dissociates into two Shockley partials. The same dissociation behaviour is observed for the interstitial-type segment shown in Fig. 9. However, when a high-energy Fe solute site is placed within the interstitial half-plane region between the two partials, as shown in Fig. 9(b1) and (b2), the partials move towards one another and recombine into a perfect dislocation, leaving almost no discernible SF ribbon.

Table 4 compares the influence of the remaining solute configurations on the separation between the two Shockley partials, together with their corresponding binding energies. In the solute-free configuration, the partial separation is approximately 13.27 Å. For the more favourable Fe site, the separation is also reduced, but by only approximately 2 Å. For H, Nb and Sn, a consistent association is observed between a smaller partial separation and a more negative binding energy. Both H and Nb generally reduce the separation relative to the solute-free configuration, whereas both tested Sn sites increase it.

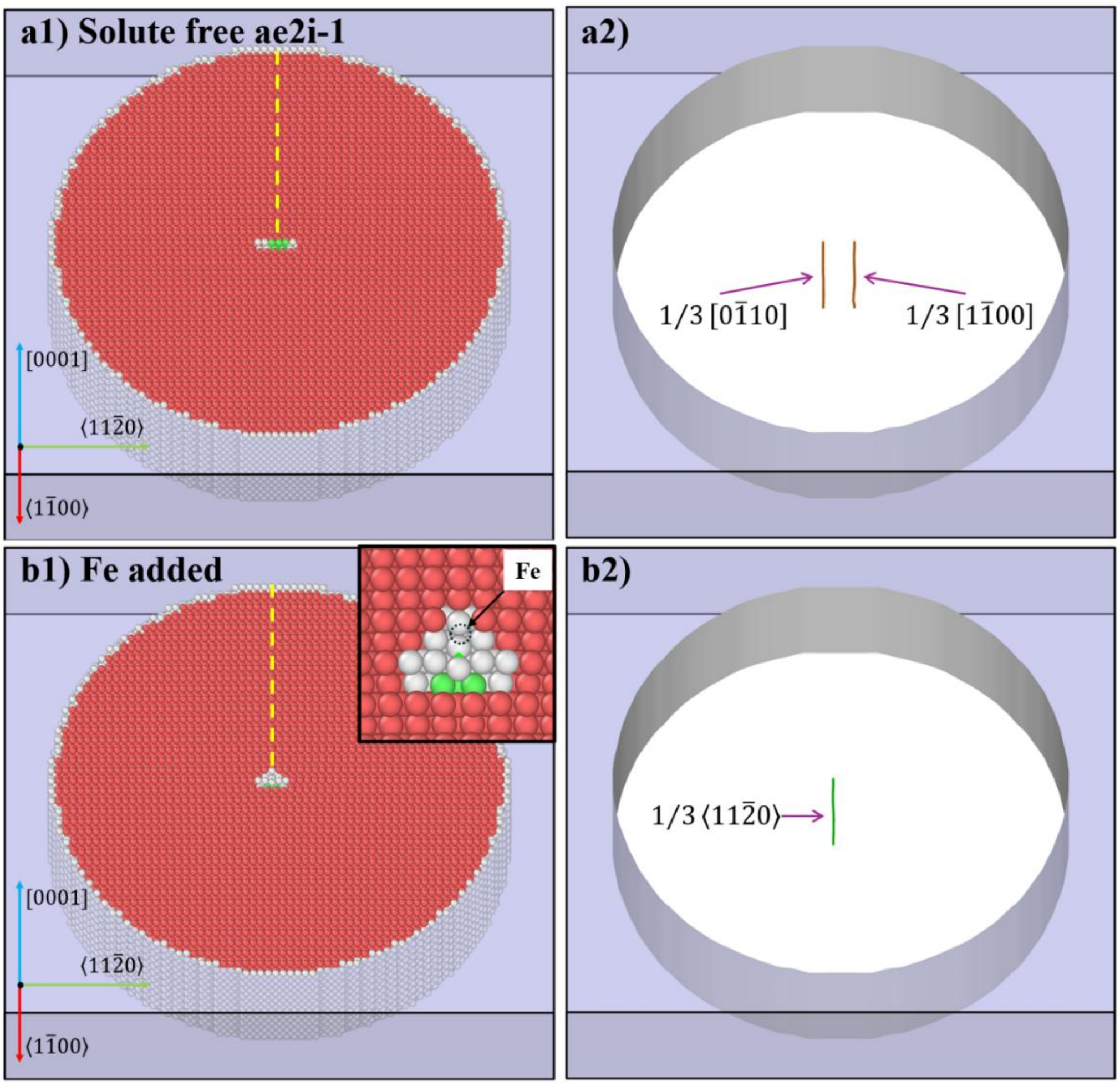

Figure 9: Case study of interstitial Fe interacting with the ae2i-1 dislocation. (1) Atomic configurations of the dislocation discs and (2) corresponding DXA-resolved dislocation-line characters for (a) the solute-free dislocation and (b) the configuration containing interstitial Fe.

Table 4: Partial-dislocation separations and QM/ML binding energies for the two selected solute configurations in ae2i-1. The solute free case separation between partials is 13.27 Å.

| Solute | High E site partial separation Å | High E site binding eV | Low E site partial separation Å | Low E site binding eV |
|---|---|---|---|---|
| Fe-int | 0 | -0.60 | 11.69 | -1.16 |
| H-int | 13.13 | -0.14 | 10.85 | -0.27 |
| Nb-sub | 12.86 | -0.05 | 12.57 | -0.56 |
| Sn-sub | 17.19 | -0.08 | 15.21 | -0.46 |

Notably, similar changes in partial separation are also observed for Fe, Nb and Sn binding to the corresponding vacancy-type segment, ae2v-1, whereas H shows the opposite trend. This contrast suggests that differences in core geometry between the vacancy- and interstitial-type dislocations influence their solute-stabilisation mechanisms.

### 3.4 Peierls barriers

The first perfect ⟨a⟩ loop is known to be glissile in Zr alloys. Here, the Peierls barriers for glide of its screw and edge components on a first-order prismatic plane are assessed. The corresponding Peierls energy profiles are presented in Fig. 10. For the edge component, the dislocation core is displaced by $1/3\langle 11\bar{2}0\rangle$. For the screw component, whose Burgers vector is parallel to the dislocation line, glide instead occurs along $[0001]$, while remaining on the same first-order prismatic plane. These glide directions are consistent with previous first-principles simulations.

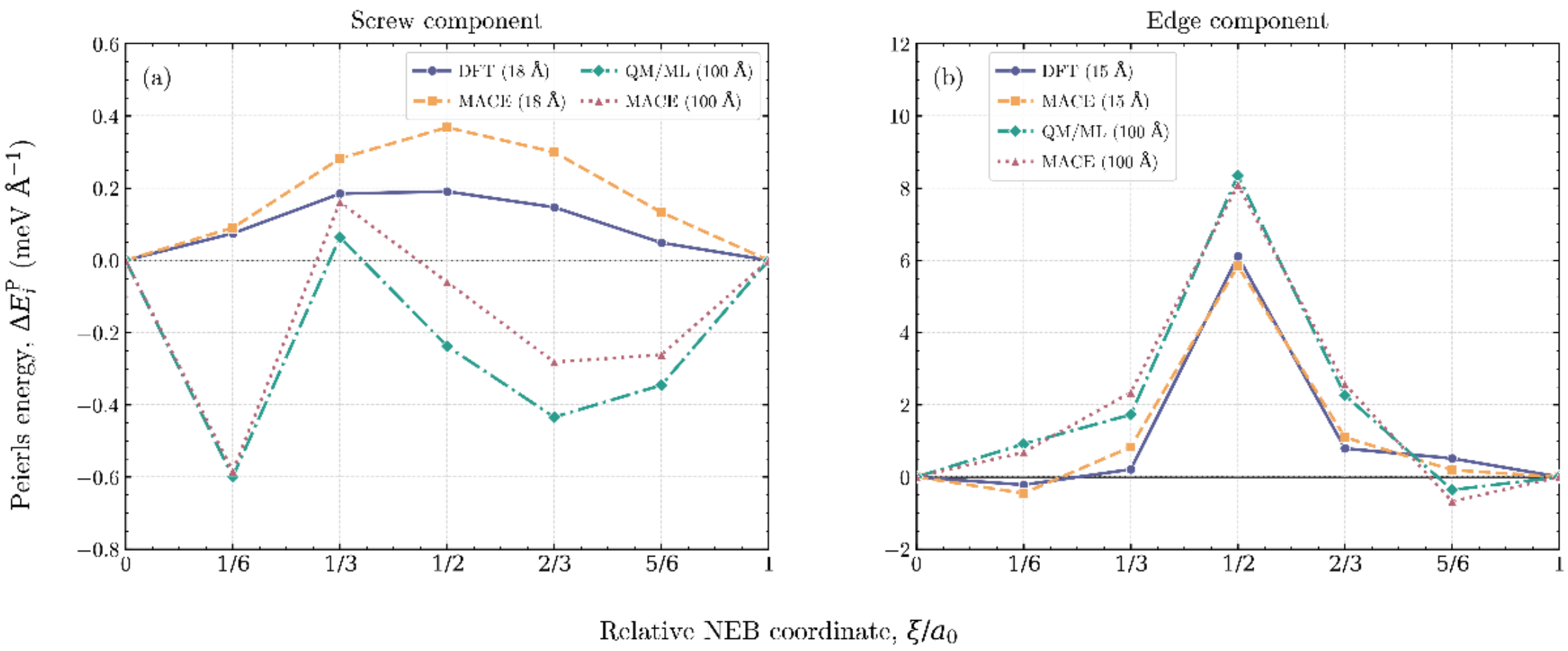

Figure 10: Peierls energy profiles for (a) the screw component and (b) the edge component of the first perfect ⟨a⟩ loop. The DFT and corresponding MLIP energy profiles were evaluated for atoms within central regions of 18 and 15 Å radius, respectively, while the QM/ML and full-disc MLIP profiles were evaluated using the converged structures from the final QM/ML step. The dislocation configurations were generated using anisotropic linear-elastic displacement fields.

The Peierls energy profiles include the full QM/ML energies, the DFT-only energies evaluated within central regions of 15 or 18 Å radius, the corresponding MACE energies for atoms within these DFT regions, and the full-disc MACE energies evaluated for the QM/ML-relaxed structures. Comparing these energy definitions and spatial regions provides further insight into methodological and finite-size effects during dislocation glide.

For the screw component, Fig. 10(a) shows that the DFT and corresponding small-region MACE profiles exhibit barriers of approximately 0.2 and 0.4 meV Å$^{-1}$, respectively. These values indicate a negligible resistance to glide and are consistent with previous DFT calculations reporting a very small activation barrier for screw-dislocation motion in Zr [19, 56]. By contrast, the full-disc QM/ML and MACE profiles differ substantially and can produce negative relative energies during glide. The similar behaviour of the full-disc MACE profile indicates that this feature originates primarily from the MLIP contribution. Although the associated energy scale remains close to zero, the incorrect sign highlights a limitation of the MLIP for resolving such small barriers.

The contribution of edge-component glide to the mobility of the first perfect ⟨a⟩ loop has received less attention. The calculated Peierls profiles show a clearer barrier of approximately 6–8.5 meV $Å^{-1}$, depending on the size of the region included in the energy evaluation. Although some variation may again arise from the MLIP contribution, the presence of an edge-glide barrier is consistently resolved across the different energy definitions.

### 3.5 Dislocation stacking fault unfaulting transition energy profile

A remaining question in understanding the geometry of ⟨a⟩ loop line segments concerns the possible transitions between loop types. The most relevant pathway is the transformation of the first faulted loop into the first perfect loop through shear on a prismatic plane. For a faulted pure-edge loop, further transformation requires the atomic half-plane to realign onto a second-order prismatic plane, giving rise to the edge component shared by the first and second perfect loops. This raises the question of whether a complete transition from the faulted loop to the first perfect loop and subsequently to the second perfect loop produces a monotonic decrease in energy, or whether distinct transition barriers are required.

For a qualitative energetic assessment, Fig. 11 summarises the transition-energy profiles, normalised by the initial stacking-fault area, for unfaulting all four faulted dislocation segments. The only clear barrier-like feature occurs for ae1i1 in the small DFT region, at approximately 1/3 of the reaction coordinate. This feature disappears when a larger region is included and is not fully reproduced by the MLIP. A barrier-like feature also appears for ae1v0 at approximately 1/6 of the reaction coordinate. However, its magnitude is only around $10^{-5}$ meV $Å^{-2}$ and should therefore not be interpreted as a physically meaningful barrier.

Beyond these small features, both the dislocation-line direction and the interstitial or vacancy character influence the transition energetics. Segments with line directions along $[0001]$ show a substantially larger energy decrease upon unfaulting than those along $\langle 11\bar{2}0\rangle$. For the $[0001]$ segments, the limited-size DFT evaluations using regions of approximately 15 Å radius give more negative transition profiles, whereas inclusion of the surrounding MLIP region in the 100 Å disc shifts the profiles towards more positive energies. The $\langle 11\bar{2}0\rangle$ segments show the opposite size dependence, with the smaller 18 Å regions producing more positive profiles than the full 100 Å discs. These trends suggest that the stability of the faulted configurations depends on both line direction and system size. The $[0001]$ faulted segments may be relatively more stable at larger sizes, while smaller systems may favour initiation of their transition. Conversely, the $\langle 11\bar{2}0\rangle$ segments appear to retain the stacking fault more readily within smaller regions.

The interstitial or vacancy character also affects the transition profile. For the $[0001]$ segments, the finite-cluster DFT profile shows a larger overall energy decrease for ae1v0 than for ae1i0, whereas the full-disc QM/ML decreases are more similar. For the $\langle 11\bar{2}0\rangle$ sthe finite-cluster DFT profile shows a barrier-like feature for ae1i1, whereas the ae1v1 profile decreases without a clearly resolved positive barrier. In addition, the MLIP consistently underestimates the magnitude of the energy variation along the NEB pathways.

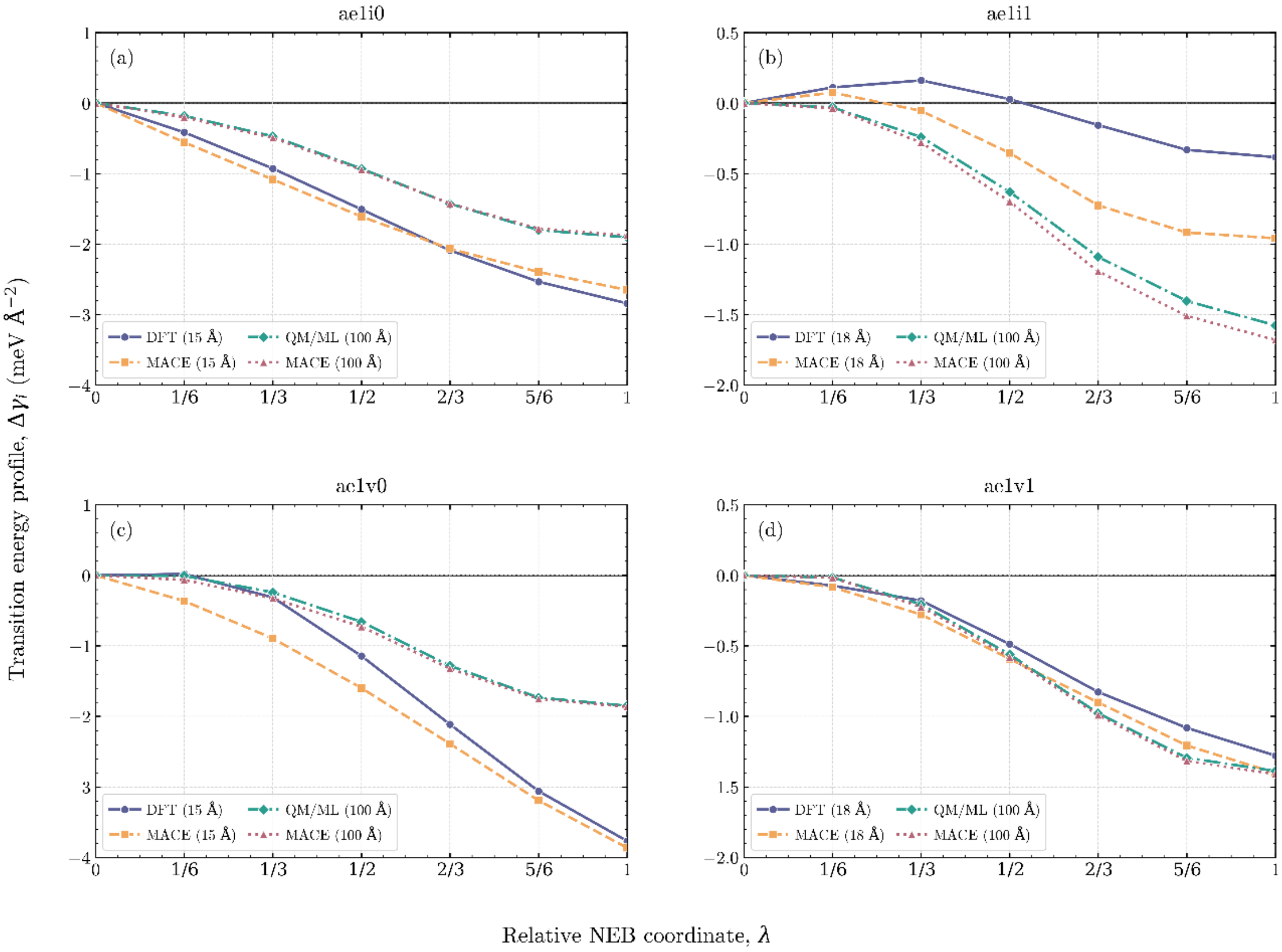

Figure 11: Unfaulting transition-energy profiles for the faulted dislocation line segments (a) ae1i0, (b) ae1i1, (c) ae1v0 and (d) ae1v1 of the first faulted ⟨a⟩ loop. The DFT and corresponding MLIP profiles were evaluated for atoms within central regions of 15 or 18 Å radius, depending on the segment, while the QM/ML and full-disc MLIP profiles were evaluated using the converged structures from the final QM/ML step.

To complete the analysis, the remaining step is to visualise the specific changes in dislocation-core geometry in the absence of solute-induced strain, so that the atomic rearrangements reflect only the intrinsic transition of the dislocation state. The ae1i0 and ae1v1 segments are taken as representative examples, to compare with the previously shown geometries in Figs. 8 and 2(b), respectively. The atomic configurations and DXA-resolved dislocation schematics at NEB coordinates 0, 1/3, 2/3 and 1 are shown in Fig. 12.

Fig. 12(a) and (b) correspond to the ae1i0 component. As the reaction coordinate progresses from 0 to 1/3, a perfect edge dislocation separates from the faulted defect mesh. As the applied shear sweeps across the faulted region from the dislocation core towards the edge, the remaining faulted defect mesh disappears, leaving a perfect edge dislocation at 2/3 with a core containing a prismatic SF ribbon oriented at 120° to the original prismatic faulted half-plane. The transition does not end at this stage, as residual stress remains to be relaxed. During the final stage, the core shifts along the blue dashed line labelled as the first prismatic plane, after which the half-plane of the new perfect edge dislocation is fully rotated onto the second prismatic plane. This corresponds to the ae1i0 to ae2i0 transition, from a faulted to a perfect loop. Referring back to the energy profiles in Fig. 11, the entire transition is energetically favourable, as indicated by the monotonic decrease in energy.

For the representative $\langle 11\bar{2}0 \rangle$ segment, the simulated route is not a faulted-edge to perfect-edge transition, but rather a faulted-edge to screw-dislocation transition, as shown in Fig. 12(c) and (d). As noted previously, the defect mesh in Fig. 12(d1) corresponds to an unidentified partial dislocation containing the prismatic stacking fault. A dissociation reaction from $1/2[1\bar{1}00]$ to a Shockley partial and an unknown partial, tentatively assigned as $1/6[1\bar{1}00]$, was proposed earlier. This missing partial is clarified by the NEB transition shown in Fig. 12(d2) to (d3), where, after shear, the unknown partial transforms into another Shockley partial. This gives the missing reaction as:

$$\frac{1}{6}[1\bar{1}00] + \frac{1}{6}[\bar{1}\bar{1}20] \rightarrow \frac{1}{3}[0\bar{1}10]$$

As a result of the transition, a clear screw character is identified. However, to distinguish it from the screw dislocation shown in Fig. 2(a), the basal second intrinsic SF ribbon that stabilised the faulted edge component remains present, suggesting an alternative core geometry for the screw component.

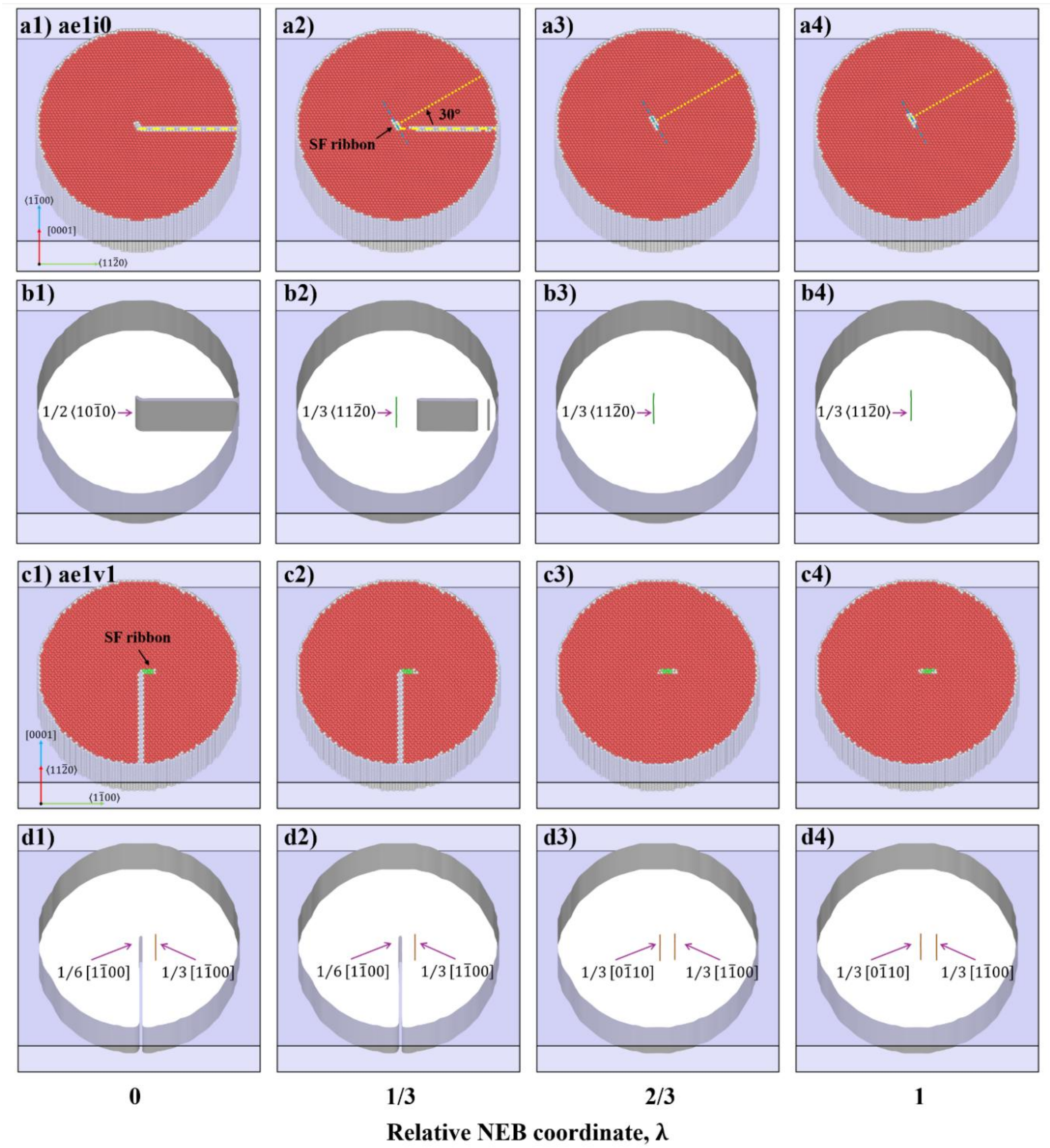

Figure 12: (a, c) Atomic configurations of the dislocation discs and (b, d) the corresponding DXA-resolved dislocation-line characters during the unfaulting processes of (a, b) the ae1i0 segment and (c, d) the ae1v1 segment. Panels (1) to (4) correspond to NEB reaction coordinates of 0, 1/3, 2/3 and 1, respectively. Yellow dashed lines indicate the atomic half-planes associated with possible edge components, while the blue dashed line marks the plane along which the partial-dislocation-induced prismatic SF ribbon extends in the ae1i0 configuration.

# 4. Discussion

## 4.1 First/second principles structural and energetic analysis of dislocations using QM/ML

Previous work demonstrated that the present QM/ML framework enables atomistic simulations that bridge the length scales accessible to DFT and those relevant to molecular dynamics [34]. By treating the dislocation core using DFT while describing the surrounding region with an MLIP, the method reduces reliance on the MLIP representation of complex defect structures and simultaneously provides means of benchmarking its performance [29, 39, 34]. In particular, solute-induced structural and energetic differences among the DFT, QM/ML and MLIP descriptions can identify limitations that may become important when the MLIP is subsequently applied to larger-scale molecular-dynamics simulations. These include simulations of radiation damage using displacement cascades and creation-relaxation algorithms, involving interactions among extended defects [49, 50].

More specifically, the deviations of the QM/ML and MLIP binding energies from the local DFT reference provide a measure of their consistency in describing solute-dislocation interactions, as shown in Fig. 7. Large deviations between the MLIP and DFT results, which are substantially reduced after QM/ML treatment, indicate limited MLIP accuracy for some interactions. The largest deviations generally occur for the interstitial solutes Fe and H, whereas the substitutional solutes Nb and Sn show smaller differences. This trend is consistent with previously reported challenges in representing interstitial defects using MLIPs [34, 51]. Interstitial Fe produces a relatively large local strain field [52], while the small size of H and its highly localised force contributions can make its atomic environment difficult to describe accurately [53, 54]. By contrast, substitutional Nb has an atomic size close to that of the Zr host and produces a comparatively small local strain field, while Sn introduces a more moderate size effect [55]. Notably, atomic size alone does not determine these interactions: local bonding, charge redistribution and the electronic and magnetic state of Fe may also contribute. Previous DFT work examined the magnetic and electronic behaviour of Fe in Zr [52], while the present binding-energy comparisons do not isolate these contributions.

Large MLIP deviations are also observed across several dislocation-loop line segments, suggesting that the core-geometry dependent energetics are not always sufficiently represented in the training data. The need for diverse defect configurations in MLIP training and validation has been highlighted previously [34]. For most configurations, these deviations are reduced by the QM/ML treatment, demonstrating that the DFT region can limit misleading MLIP contributions within the dislocation core. At the same time, the surrounding MLIP region retains contributions from longer-range elastic and solute-strain interactions that extend beyond the finite DFT region. The remaining difference between the QM/ML and local DFT energies may therefore contain both genuine long-range contributions and residual MLIP error, consistent with the interpretation established in our previous methodological work [34].

A major extension of the framework in the present work is the implementation of the QM/ML-NEB method for evaluating transition barriers. This enables comparison with previous first-principles studies of dislocation migration and Peierls barriers in Zr [19, 56] and provides access to both dislocation glide and faulted-to-perfect transition pathways. A key motivation for introducing this capability is the overprediction of the prismatic stacking-fault energy by the current MLIP, as identified in the preceding benchmark. This error reduces the predicted stability of dislocations containing the prismatic fault. During unrestricted MLIP optimisation, the faulted dislocations can therefore transform spontaneously into perfect configurations. A less restricted convergence criterion was required to retain these faulted structures during the MLIP pre-relaxation. Under comparable relaxation conditions, the QM/ML treatment preserves the prismatic fault, indicating that the spontaneous transition is primarily associated with the MLIP description of the faulted core.

A related distinction is observed when solutes are introduced. The QM/ML calculations show that the combined effects of solute chemistry and local strain can modify the dislocation-core geometry and, in some cases, separate a perfect-dislocation component from the faulted configuration. Solute-induced changes in dislocation-core geometry and partial separation have also been reported in previous atomistic studies [19, 56, 57, 58]. However, the corresponding MLIP-only calculations can instead predict complete unfaulting and elimination of the prismatic fault. This differs from the localised, partial reconstruction obtained when the core is treated using DFT. Comparing transition-energy profiles calculated using DFT-containing and MLIP-dominated energy definitions is therefore necessary to distinguish physically supported core reconstruction from MLIP-driven unfaulting. For the transitions examined here, the MLIP profiles generally lie below the corresponding DFT-containing profiles, consistent with the reduced stability of the faulted structures predicted by the MLIP.

## 4.2 Interactions between solutes and ⟨a⟩ loop line segments in Zr alloys

The central question addressed in this work concerns the interactions between solutes and irradiation-induced dislocation-loop line segments in Zr alloys. For ⟨a⟩ loops, the present analysis covers the three possible loop configurations, including faulted and perfect loops, interstitial- and vacancy-type loops, and edge and screw components. The representative solutes occupy their preferred states as in perfect Zr, with Fe and H treated as interstitial solutes and Nb and Sn as substitutional solutes. Combining the results in Figs. 3 to 6 reveals systematic solute preferences among the loop configurations, while the representative case studies illustrate how these interactions can modify the dislocation-core structure.

Previous DFT studies examined solute interactions with stacking faults in Zr alloys to assess a possible Suzuki effect, whereby solute segregation reduces the stacking-fault energy and stabilises the faulted structure [17, 59]. These calculations identified pronounced attraction between interstitial Fe and the prismatic stacking fault, limited attraction or repulsion for substitutional Nb, and strong repulsion for substitutional Sn [17]. Similar trends remain when the more complex dislocation-core geometries are considered here. Fe preferentially binds to the faulted-loop segments rather than to the perfect-loop segments, whereas Nb and Sn can be repelled by the faulted segments and therefore show a relative preference for the perfect loops. Interstitial H also exhibits a stronger preference for the faulted dislocations, although H was not included in the previous stacking-fault and solute-interaction study.

Previous work in Ti, another Group IV metal, showed that a prismatic stacking fault can produce a distinct octahedral-like interstitial environment that stabilises interstitial solutes such as Fe [60]. In the present Zr calculations, the evolution of SF ribbons between partials and the extension of the faulted core onto adjoining prismatic planes increase the range of locally distorted environments that may accommodate interstitial solutes. This provides a possible structural explanation for the stronger Fe and H binding to the faulted-loop segments. The comparison with Ti is further supported by previous DFT analysis of the magnetic and electronic stabilisation of interstitial and substitutional Fe in Zr, Ti and Hf [52]. Although the detailed energies remain host dependent, these studies indicate that related Fe-stabilisation mechanisms can occur across the Group IV transition metals.

The calculated loop preferences can also be considered in relation to experimentally observed radiation-induced segregation. Previous TEM and atom-probe tomography studies reported planar solute distributions associated with ⟨a⟩-loop arrays, with Fe enriched near the loop planes and Sn- and Nb-rich regions located between them [9]. DFT studies of small solute clusters have further examined Fe-Sn and Fe-Nb interactions over a range of cluster sizes and configurations [55]. These calculations identified short-range repulsion between Sn and Fe, and attraction between Nb and Fe, partly attributed to the magnetic and electronic behaviour of Fe in Zr [52, 55]. The predicted Sn-Fe repulsion is consistent with the experimentally observed Sn-Fe anticorrelation, whereas the calculated Nb-Fe attraction does not directly explain the corresponding experimental Nb-Fe anticorrelation.

When combined with differences in solute mobility and loop preference, the present results suggest a possible sequence for the development of these segregation patterns. Interstitial Fe is expected to diffuse more rapidly than substitutional Sn and Nb in the Zr lattice [61]. If ⟨a⟩ loops develop sequentially from the first faulted loop to the first perfect loop and subsequently to the second perfect loop [12], Fe may be trapped during the early faulted stage and remain bound after the transition to a perfect configuration. The slower-diffusing Sn solutes may subsequently be attracted to the perfect-loop segments, while their distribution is constrained by repulsion from the Fe already present. For Nb, the interpretation is less direct. Its comparatively weak interactions within some Fe-Sn-Nb cluster configurations may allow it to associate with Sn-rich regions, despite the calculated attraction between Nb and Fe being short range [55].

The segregation behaviour may also be influenced by interactions involving H. Repulsion between Sn and interstitial solutes is not limited to Fe, as Sn can also be repelled by H-containing configurations [62]. Conversely, Nb can attract and trap H in the Zr lattice [62]. These competing interactions provide an additional possible contribution to the relative distributions of Fe, Sn and Nb around irradiated ⟨a⟩-loop arrays. Overall, the present binding energies provide a basis for these proposed mechanisms, although the complete segregation sequence will also depend on solute diffusion, irradiation flux, defect production and the kinetics of loop evolution.

### 4.3 Insights into the geometry and transitions of ⟨a⟩ loop line segments

A remaining aspect of the atomic-scale behaviour of loops concerns their dislocation-core geometries and transition mechanisms. As experimentally observed loops are commonly reported to have perfect Burgers vectors [9, 11, 15, 16], their stabilisation may depend partly on dissociation into partial dislocations. In the present results, Shockley partials with Burgers vectors belonging to the $1/3\langle 1\bar{1}00\rangle$ family contribute to the core structures of both perfect and faulted loop segments. Perfect edge dislocations generally dissociate into two Shockley partials separated by a second intrinsic basal SF ribbon, consistent with previous atomistic simulations [19]. Among the possible basal stacking faults in Zr, this second intrinsic fault has received comparatively limited attention [12, 17]. It can be formed by shear without the insertion or removal of an atomic plane, unlike the first intrinsic and extrinsic basal faults that are commonly associated with ⟨c⟩-component loops [14].

For the faulted dislocations, particularly those with line directions along $\langle 11\bar{2}0\rangle$, the core contains both a Shockley partial and a half-Shockley partial with a Burgers vector belonging to the $1/6\langle 1\bar{1}00\rangle$ family. A similar half-Shockley partial has been observed by TEM in Zr and associated with the transformation sequence from hcp Zr through an intermediate structure towards an fcc-like configuration [63]. In the present calculations, this partial is not identified directly by DXA because the transformed local crystal structure and associated stacking fault prevent complete dislocation-line reconstruction. However, Burgers-vector conservation, the atomistic core geometry and the QM/ML-NEB transition pathway collectively support its assignment.

The transition calculations also provide insight into the half-plane rotation of faulted edge dislocations with line directions along $[0001]$. Following unfaulting, the atomic half-plane of the resulting perfect dislocation is oriented at 30° to the removed half-plane of the original faulted configuration. The corresponding perfect $[0001]$ edge dislocation can be interpreted as dissociating into two Shockley partials associated with half-planes rotated by 30° in opposite directions. One partial retains a half-plane aligned with the original faulted plane, while the second forms on a plane oriented at 60° to it, thereby accommodating the remaining strain and stabilising the perfect-dislocation core. Although DXA does not resolve this dissociation directly, the prismatic SF ribbon within the core is consistent with the presence of the two partials and provides a possible explanation for the limited DXA resolution.

The energy profiles for the $[0001]$ and $\langle 11\bar{2}0\rangle$ faulted segments also show contrasting contributions from the local core and the surrounding atomistic region. For the $\langle 11\bar{2}0\rangle$ segments, the limited-size DFT only profiles are relatively more positive than the corresponding full-disc profiles, whereas the $[0001]$ segments show the opposite trend. Because these curves represent different spatial contributions to the same transition rather than independently simulated loop sizes, they do not directly establish a physical loop-size dependence. Instead, they indicate that the local core and longer-range elastic region contribute differently to unfaulting for the two line directions. The more favourable local transition profile of the $[0001]$ segments suggests that these regions may provide preferential initiation sites for unfaulting, while the subsequent propagation of shear around a complete loop would depend on the energetics of the remaining line segments.

Finally, perfect loops in Zr alloys are not always circular and can adopt elliptical morphologies, particularly for vacancy-type loops [11, 12, 16]. This behaviour has previously been attributed partly to anisotropic vacancy and self-interstitial diffusion [64, 65]. The present Peierls-barrier calculations suggest an additional mechanism that may contribute to this anisotropy. The ideal screw component shows a negligible glide barrier within the numerical resolution of the calculations, consistent with previous DFT results [19, 56], whereas the edge component exhibits a barrier of approximately 6–8.5 meV $Å^{-1}$ across the energy definitions shown in Fig. 10(b). This contrast indicates that the mobility of individual loop segments can depend strongly on their dislocation character and may therefore contribute to anisotropic loop-shape evolution.

An important distinction must be made between an ideal perfect loop formed directly through conservative shear and a perfect loop produced by unfaulting an interstitial- or vacancy-type faulted loop. The as1 screw segment examined in the Peierls calculation represents the former case. It was generated using an anisotropic elastic displacement field and contains no explicit excess or deficit of atoms. Its negligible Peierls barrier therefore describes the mobility of an idealised, locally conservative screw dislocation. By contrast, shear-driven unfaulting of an interstitial- or vacancy-type faulted loop changes its fault structure and local dislocation character without removing the overall excess or deficit of atoms. The resulting perfect loop consequently retains the interstitial- or vacancy-type character of its precursor.

For the $\langle 11\bar{2}0\rangle$ faulted segments, unfaulting shears the prismatic fault and produces a nominally screw-character segment with the same line and glide directions as the ideal as1 screw component. However, the retained basal SF ribbon and associated partial-dislocation structure show that the transformed core is not equivalent to the ideal screw core. The ribbon reflects the structural history inherited from the faulted loop and may introduce an additional non-conservative or edge-like contribution that restricts glide. The calculations therefore support a pathway by which experimentally observed interstitial- and vacancy-type perfect loops can form through unfaulting of faulted precursors while retaining core features that contribute to their stability.

This inherited core structure may also help explain why the nominal screw components of a perfect loop do not simply glide towards one another and annihilate despite the negligible barrier calculated for the ideal screw configuration. Their motion remains coupled to the retained SF ribbon, the adjoining segments and the elastic constraints of the complete closed loop. Solutes can further modify this behaviour. Fe, for example, can reduce the width of the SF ribbon and promote partial recombination, potentially changing the mobility of the transformed segment. The combined effects of dislocation character, transformation history, retained fault structure and solute-induced core rearrangement may therefore contribute to the development of elliptical loops. Direct confirmation would require simulations of the coupled motion and morphological evolution of complete loops.

## 5. Conclusions

1. A hybrid QM/ML framework was applied to the core structures, solute interactions, Peierls barriers and unfaulting pathways of ⟨a⟩-loop dislocation segments in α-Zr. DFT treatment of the core reduces dependence on the MLIP while retaining the surrounding elastic field.
2. Nine line segments representing the three possible ⟨a⟩-loop configurations were constructed and relaxed. Their atomistic structures demonstrate the importance of Shockley partials in both perfect and faulted loops.
3. Atomistic, Burgers-vector and QM/ML-NEB analyses support a half-Shockley partial in the faulted $\langle 11\bar{2}0\rangle$ segments. Together with a Shockley partial, it bounds a second intrinsic basal SF ribbon. Basal and prismatic SF ribbons contribute more generally to core stabilisation.
4. Interstitial Fe and H bind preferentially to faulted-loop segments, whereas substitutional Sn and Nb show a relative preference for perfect-loop segments. These trends provide a possible thermodynamic basis for planar solute segregation around irradiated ⟨a⟩-loop arrays.
5. Solute chemistry and local strain can displace or recombine partials, change SF-ribbon widths and locally promote unfaulting. Solute stabilisation therefore involves both direct binding and dislocation-core reconstruction.
6. The ideal conservative screw component has a negligible Peierls barrier, whereas the edge component of the first perfect ⟨a⟩ loop has a substantially higher barrier of approximately 6–8.5 meV $Å^{-1}$. Screw-character segments inherited through unfaulting retain additional SF-ribbon structure, which may restrict their mobility. These differences may contribute to anisotropic and elliptical loop evolution.
7. QM/ML-NEB reveals line-direction-dependent unfaulting pathways and supports partial reactions that are not fully resolved by DXA. MLIP overprediction of the metastable prismatic stacking-fault energy can artificially favour unfaulting, demonstrating the importance of DFT treatment for reliable fault stability and transition energetics.

## Acknowledgements

We are grateful to the UK Materials and Molecular Modelling Hub for computational resources, which is partially funded by EPSRC (EP/T022213/1, EP/W032260/1 and EP/P020194/1). We acknowledge computational resources and support provided by the Imperial College Research Computing Service (http://doi.org/10.14469/hpc/2232). ChatGPT was used to assist with grammatical polishing, improving readability and providing guidance during the development of the Python code.

## Data availability

All data generated or analysed during this study are available from the corresponding author on reasonable request.

## Code availability

The code supporting the findings of this study is publicly available at https://github.com/antony0429/Hybrid-QM-ML-framework.

## Supporting Information

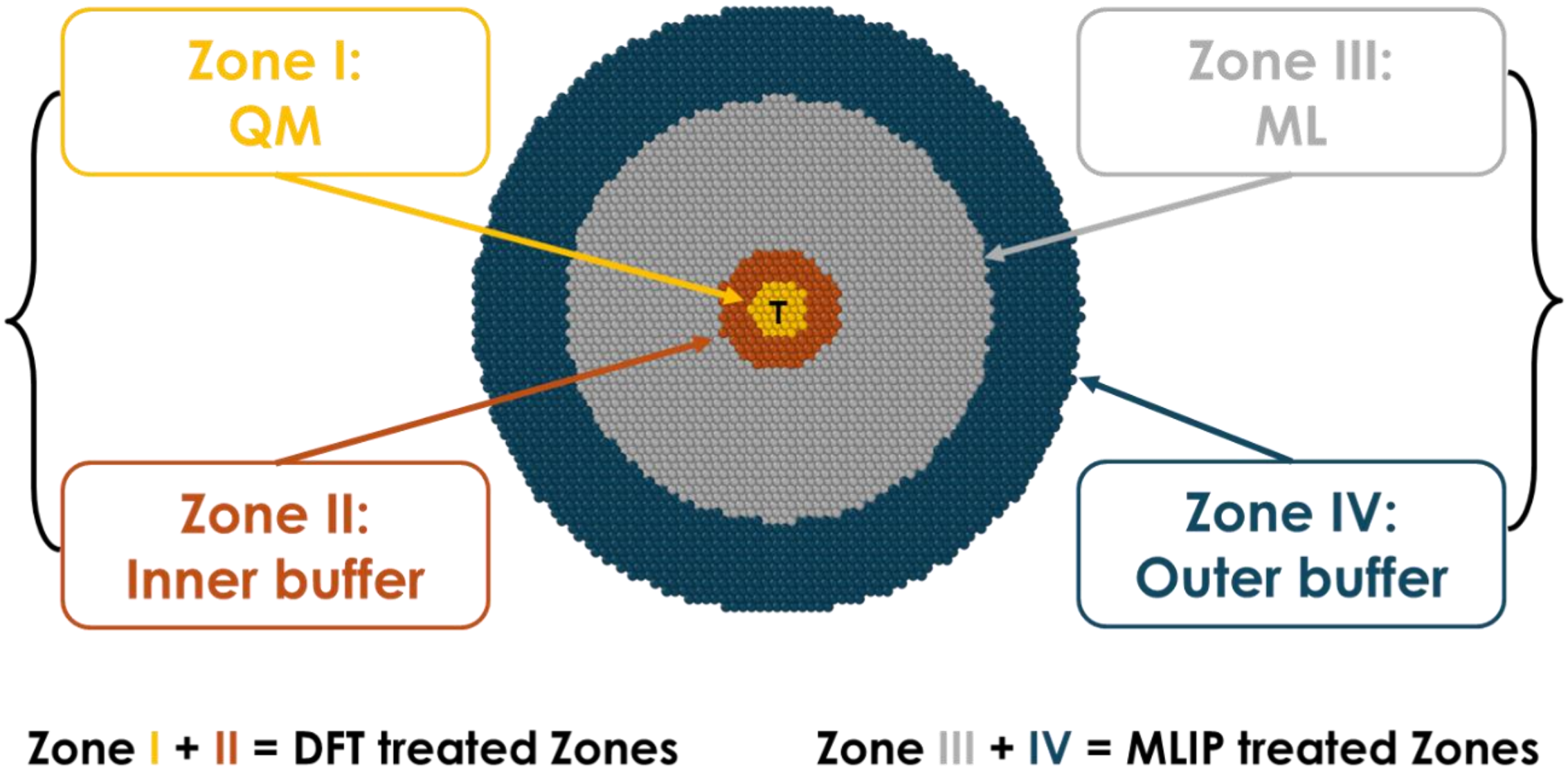

Figure S1: Four region disk shaped QM/ML model used in this work. Zone I is the QM region containing the chemically sensitive defect core and solutes, Zone II is the inner buffer included in the DFT calculation, Zone III is the MLIP embedding region, and Zone IV is the constrained outer buffer used to reduce boundary artefacts. Reproduced from authors' own previous work [34]

Elastic constant predictions from the universal MACE medium-mpa-0 potential for Zr, compared with DFT.

| Method | C11 GPa | C12 GPa | C13 GPa | C33 GPa | C44 GPa | C66 GPa |
|---|---|---|---|---|---|---|
| **DFT** | **141.1** | **67.6** | **64.3** | **166.9** | **25.8** | **36.8** |
| MACE | 144.3 | 65.8 | 69.0 | 160.8 | 35.8 | 39.2 |

Defect formation energy predictions from the universal MACE medium-mpa-0 potential for Zr, compared with DFT. With s for substitutional solutes and i for interstitial solutes.

| Method | Vacancy eV | Sn-s eV | Nb-s eV | Fe-i eV | H-i eV |
|---|---|---|---|---|---|
| DFT | 2.06 | -1.16 | 0.60 | 0.93 | -0.45 |
| MACE | 2.25 | -1.17 | 0.52 | 1.01 | -0.55 |